\documentclass[aps,prb,twocolumn,letterpaper]{revtex4}

\usepackage{amsmath,amssymb,amsfonts,upgreek}
\usepackage{bm}
\usepackage{booktabs}
\usepackage{dcolumn}
\usepackage{color}
\usepackage{graphicx}
\usepackage{setspace}
\usepackage{multirow}

\begin{document}

\title{Effect of Hubbard $U$ on the construction of low energy
  Hamiltonians for LaMnO$_3$ via maximally localized Wannier
  functions}

\date{\today}

\author{Roman Kov\'a\v{c}ik}
\affiliation{School of Physics, Trinity College Dublin, Dublin 2, Ireland}
\email{kovacikr@tcd.ie}
\author{Claude Ederer}
\affiliation{School of Physics, Trinity College Dublin, Dublin 2, Ireland}

\begin{abstract}
We use maximally localized Wannier functions to construct
tight-binding (TB) parameterizations for the $e_g$ bands of LaMnO$_3$
based on first principles electronic structure calculations. We
compare two different ways to represent the relevant bands around the
Fermi level: i) a \emph{$d$-$p$ model} that includes atomic-like
orbitals corresponding to both Mn($d$) and O($p$) states in the TB
basis, and ii) an \emph{effective $e_g$ model} that includes only two
$e_g$-like Wannier functions per Mn site. We first establish the
effect of the Jahn-Teller distortion within the $d$-$p$ model, and
then compare the TB representations for both models obtained from
GGA+$U$ calculations with different values of the Hubbard parameter
$U$. We find that in the case of the $d$-$p$ model the TB parameters
are rather independent on the specific value of $U$, if compared with
the mean-field approximation of an appropriate multi-band Hubbard
Hamiltonian. In contrast, the $U$ dependence of the TB parameters for
the effective $e_g$ model cannot easily be related to a corresponding
mean-field Hubbard model, and therefore these parameters depend
critically on the specific value of $U$, and more generally on the
specific exchange-correlation functional, used in the electronic
structure calculation.
\end{abstract}


\maketitle

\section{Introduction}

The construction of realistic low-energy Hamiltonians based on first
principles electronic structure calculations is an important tool for
the investigation of correlated electron systems, such as
e.g. cuprates, manganites, or other complex transition metal (TM)
oxides.  While much of the general physics determining the diverse
properties of these materials can be understood in terms of simplified
models,\cite{Dagotto:1994,Imada/Fujimori/Tokura:1998,Dagotto/Hotta/Moreo:2001}
the corresponding model parameters are generally unknown and have to
be adjusted to fit experimental data. Alternatively, electronic
structure calculations based on density functional theory
(DFT)~\cite{Hohenberg/Kohn:1964,Kohn/Sham:1965} can be used to
determine these
parameters,\cite{Gunnarsson_et_al:1989,Hybertsen/Schlueter/Christensen:1989}
which in turn allows for a quantitative evaluation of the underlying
model assumptions and the construction of realistic model Hamiltonians
with all parameters determined \emph{ab
  initio}.\cite{Ederer/Lin/Millis:2007,Kovacik/Ederer:2010}

Model Hamiltonians for correlated electron systems are typically
formulated within a tight-binding (TB) picture that involves only a
small number of electronic states localized on certain atoms (e.g. the
``$d$-states'' of the TM atoms within a TM
oxide).\cite{Dagotto:1994,Imada/Fujimori/Tokura:1998,Dagotto/Hotta/Moreo:2001}
Ideally, these states give rise to an isolated group of bands around
the Fermi energy, which then determines the low energy behavior of the
system. A corresponding TB representation can in principle be obtained
by constructing Wannier functions from the Kohn-Sham Bloch states
calculated within
DFT.\cite{Mueller_et_al:1998,Ku_et_al:2002,Zurek/Jepsen/Andersen:2005,Lechermann_et_al:2006,Solovyev:2006,Kovacik/Ederer:2010}

In the case of model Hamiltonians that contain an explicit
electron-electron interaction, typically in the form of a local
Hubbard term with interaction parameter $U$, the DFT band-structure
can either be viewed as mean-field approximation to this interacting
model, or as representative for the ``non-interacting'' case,
i.e. corresponding to $U=0$ in the model Hamiltonian. While it is
often convenient to consider the electronic structure calculated
within either the local density approximation
(LDA)~\cite{Hohenberg/Kohn:1964,Perdew/Zunger:1981} or the generalized
gradient approximation (GGA)~\cite{1996_perdew} as essentially
non-interacting, this is probably not a good assumption if the
electronic structure is calculated using more advanced exchange
correlation functionals such as (LDA/GGA)+$U$ or hybrid
functionals.\cite{Liechtenstein/Anisimov/Zaanen:1995,1998_dudarev,Becke:1993}

In this work we use maximally localized Wannier functions
(MLWFs)~\cite{1997_marzari,2001_souza,Mostofi_et_al:2008} to obtain TB
parameterizations for the important case of LaMnO$_3$, the parent
material of the colossal magneto-resistive manganites, and a prototype
system for correlated electron
physics.\cite{Coey/Viret/Molnar:1999,Dagotto/Hotta/Moreo:2001} We
calculate the electronic structure of LaMnO$_3$ using the GGA+$U$
method and different values for the Hubbard parameter $U$. We then
compare two different ways to represent the relevant bands around the
Fermi level: i) a \emph{$d$-$p$ model} that includes atomic-like
orbitals corresponding to both Mn($d$) and O($p$) states in the TB
basis, and ii) an \emph{effective $e_g$ model} that includes only two
$e_g$-like Wannier functions per Mn site. In particular, we analyze
the $U$ dependence of the two different TB parameterizations and
relate this to commonly used model Hamiltonians for manganites.

We find that for the $d$-$p$ parameterization, the effect of $U$ is
mostly local, leading to $U$ dependent shifts of the on-site energies
and an increase in the Jahn-Teller (JT) splitting, whereas the
corresponding hopping amplitudes are only weakly affected by the value
of $U$. Thus, the $U$ dependence of the $d$-$p$ parameters closely
resembles the $U$ dependence of a corresponding multi-band Hubbard
Hamiltonian in mean-field approximation. On the other hand, the change
of on-site energies and JT splitting for the effective $e_g$ MLWFs,
calculated for different values of $U$, are distinctly different from
a corresponding mean-field model Hamiltonian. In addition, there is
also a strong $U$ dependence of the effective $e_g$ hopping
amplitudes, which is due to electronic degrees of freedom that are
excluded from the TB basis. This indicates that while the TB
parameterization for the $d$-$p$ model is fairly robust with respect
to a variation of $U$, for the case of the effective $e_g$ basis an
appropriate choice of $U$ in the GGA+$U$ calculation is crucial. Our
results demonstrate the simple fact that the transferability of a
specific TB parameterization is generally increased if more electronic
degrees of freedom are included in the model description.

This paper is organized as follows. In the next section we first
summarize the most important aspects of typical model Hamiltonians for
manganites, emphasizing in particular the role of the JT distortion
(Sec.~\ref{ssec:met_jt}), before introducing MLWFs
(Sec.~\ref{ssec:mlwf}) and describing the technical details of our
calculations (Sec.~\ref{ss:comp-details}). In Secs.~\ref{ss:cubic} and
\ref{ss:dp-jt} we first present the main features of the $d$-$p$ model
based on the MLWFs calculated for $U=0$~eV, and then clarify the
effect of the JT distortion on the corresponding TB
parameterization. An analogous analysis for the effective $e_g$ model
parameterization has recently been presented in
Ref.~\onlinecite{Kovacik/Ederer:2010}. The effect of varying the
Hubbard $U$ in the GGA+$U$ calculation on the MLWF parameters of the
two different models is presented in Sec.~\ref{ss:dp-jt-u} for the
$d$-$p$ model and in Sec.~\ref{ss:dd-jt-u} for the effective $e_g$
model. Finally, Sec.~\ref{sec:summary} summarizes our main
conclusions, while further details about the $d$-$p$ TB
parameterization can be found in Appendix~\ref{sec:appendix}.

\section{Method and theoretical background}\label{s:met}

\subsection{Jahn-Teller distortion in LaMnO$_3$}\label{ssec:met_jt}

LaMnO$_{3}$ crystallizes in an orthorhombically distorted perovskite
structure with $Pbnm$ space group symmetry.\cite{Elemans_et_al:1971}
The distortion relative to the cubic perovskite structure can be
decomposed into three components:\cite{Ederer/Lin/Millis:2007} a
staggered JT distortion of the MnO$_{6}$ octahedra, alternating tilts
and rotations of these octahedra around the cubic axes (GdFeO$_{3}$
distortion), and an orthorhombic strain of the unit cell. The
electronic structure of LaMnO$_3$ around the Fermi energy is dominated
by Mn $3d$ states, which are split by the octahedral crystal field
into lower-lying $t_{2g}$ and higher-lying $e_{g}$
states.\cite{Pickett/Singh:1996,Satpathy/Popovic/Vukajlovic:1996,Sawada_et_al:1997}
The formal $d^4$ occupation of the Mn$^{3+}$ cation leads to a high
spin configuration with filled majority spin $t_{2g}$ states,
half-filled majority spin $e_{g}$ states, and empty minority spin
states. The low energy behavior of LaMnO$_{3}$ is therefore governed
by the partially filled majority spin bands with predominant $e_{g}$
character.

Motivated by this, the complex phenomenology observed in manganites is
often modeled within an effective TB model that involves only two
$e_g$ orbitals per Mn site. Electrons can then hop between the two
$e_g$ levels on neighboring sites, and can interact with each other
through a Hubbard-type electron-electron interaction, with the
$t_{2g}$ ``core-spins'' through a Hund's rule interaction, and with
the local JT distortion through a crystal field splitting (see
e.g. Ref.~\onlinecite{Dagotto/Hotta/Moreo:2001}). It is usually
understood that the corresponding ``$e_g$ orbitals'' are spatially
extended Wannier orbitals that result from hybridization between
atomic-like Mn($e_g$) orbitals and the $p$ orbitals of the surrounding
oxygen ligands.

The effect of the JT distortion on this $e_g$ manifold is typically
expressed via a local crystal-field splitting of the form:
\begin{equation}
  \label{eq:jt}
  \hat{H}_\text{JT} = - \lambda \sum_{\mathbf{R},\sigma,a,b,i}
  \hat{c}^\dagger_{a\mathbf{R}\sigma} Q^i_\mathbf{R}
  \tau^i_{ab} 
  \hat{c}_{b\mathbf{R}\sigma}\ .
\end{equation}
Here, $\lambda$ is the JT coupling strength, $\bm{\tau}_{ab}$ are the
usual Pauli matrices, $\hat{c}_{b\mathbf{R}\sigma}$ is the
annihilation operator corresponding to orbital $b$ with spin $\sigma$
at site $\mathbf{R}$, and the JT distortion of the oxygen octahedron
surrounding site $\mathbf{R}$ is described by specific modes
$Q^i_\mathbf{R}$ ($i=x,z$).\cite{Dagotto/Hotta/Moreo:2001}

\begin{figure}
\centerline{
\includegraphics[width=0.8\columnwidth]{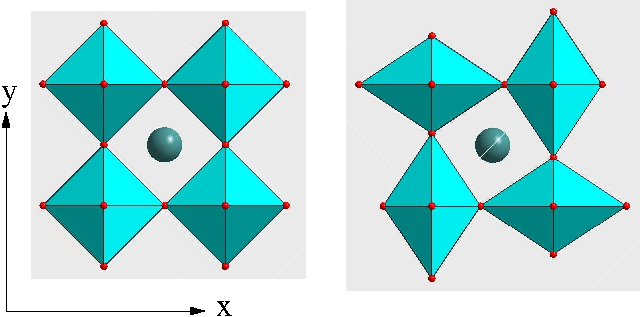}}
\caption{(Color online) Left: ideal cubic perovskite structure viewed
  along the $z$ axis. Right: staggered $Q^x$-type JT distortion within
  the $x$-$y$ plane. Oxygen anions are shown as small (red) spheres,
  the La cation as large (green) sphere. Mn cations (not shown) are
  situated in the middle of each oxygen octahedron.}
\label{fig:JT-dist}
\end{figure}

In the following we only consider a staggered JT distortion of the
form $Q^x_\mathbf{R} = - Q^x_{\mathbf{R}'}$, where $\mathbf{R}$ and
$\mathbf{R}'$ correspond to nearest neighbor sites within the $x$-$y$
plane (see Fig.~\ref{fig:JT-dist}). Thereby:
\begin{equation}
  Q^x_\mathbf{R} = \frac{1}{\sqrt{2}} \left(d^x_{\mathbf{R}} -
  d^y_\mathbf{R} \right) \ ,
\end{equation}
with $d^x_\mathbf{R}$ and $d^y_\mathbf{R}$ being the Mn-O distances
along the $x$ and $y$ directions, respectively, corresponding to the
oxygen octahedron surrounding the Mn at site $\mathbf{R}$. We note
that it has been shown in Ref.~\onlinecite{Ederer/Lin/Millis:2007}
that this component of the JT distortion has the most pronounced
effect on the electronic structure of LaMnO$_3$. Furthermore, in
Ref.~\onlinecite{Kovacik/Ederer:2010} we showed that the effect of the
various structural distortions on the calculated model parameters can
be analyzed separately, since they are to a great extent independent
of each other.

\subsection{Maximally localized Wannier functions} \label{ssec:mlwf}

A set of $N$ localized Wannier functions $\lvert w_{n\mathbf{T}}
\rangle$ corresponding to a group of $N$ bands that are described by
delocalized Bloch states $|\psi_{m\mathbf{k}}\rangle$, is defined by
the following transformation:
\begin{equation}
  \label{eq:mlwf}
  \lvert{w_{n\mathbf{T}}}\rangle = \frac{V}{\left({2\pi}\right)^{3}}
  \int_{\mathrm{BZ}} \mathrm{d}\mathbf{k} \left[{\sum_{m=1}^{N}
      U_{mn}^{\left(\mathbf{k}\right)}
      \lvert{\psi_{m\mathbf{k}}}\rangle} \right]
  \mathrm{e}^{-\mathrm{i}\mathbf{k}\cdot\mathbf{T}}
  \,.
\end{equation}
Thereby, $\mathbf{T}$ is the lattice vector of the unit cell
associated with the Wannier function, $m$ is a band index,
$\mathbf{k}$ is the wave-vector of the Bloch function, and the
integration is performed over the first Brillouin zone (BZ) of the
lattice. Different choices for the unitary matrix
$\mathbf{U}^{(\mathbf{k})}$ lead to different Wannier functions, which
are thus not uniquely defined by Eq.~(\ref{eq:mlwf}). A unique set of
\emph{maximally localized Wannier functions} (MLWFs) can be generated
by minimizing the total quadratic spread of the Wannier
orbitals.\cite{1997_marzari}

Once the transformation matrices $\mathbf{U}^{(\mathbf{k})}$ are
determined, a TB representation of the Hamiltonian in the MLWF basis
is obtained:
\begin{equation}
  \label{eq:tbh}
  \hat{H} = \sum_{\mathbf{T}, \Delta\mathbf{T}} h_{nm}^{\Delta\mathbf{T}}
  \, \hat{c}^\dagger_{n\mathbf{T}+\Delta\mathbf{T}} \hat{c}_{m\mathbf{T}}
  \ + \text{h.c.} \ ,
\end{equation}
with
\begin{equation}
  \label{eq:hr}
  h^\mathbf{T}_{nm} = \frac{V}{(2\pi)^3} \int_\text{BZ} \mathrm{d}\mathbf{k}
  \left[
  \sum_{l} \left(U^{(\mathbf{k})}_{ln}\right)^* \epsilon_{l\mathbf{k}} \, U^{(\mathbf{k})}_{lm} \right]
\mathrm{e}^{-\mathrm{i}\mathbf{k}\cdot\mathbf{T}}  \,.
\end{equation}
Here, $\epsilon_{l\mathbf{k}}$ is the eigenvalue corresponding to
Bloch function $\lvert \psi_{l\mathbf{k}} \rangle$. For cases where
the bands of interest do not form an isolated set of bands but are
entangled with other bands, a two step procedure for obtaining the
unitary transformation matrices (which in this case are typically
rectangular) is employed.\cite{2001_souza}

We note that $\mathbf{T}$ and $\Delta \mathbf{T}$ in
Eqs.~(\ref{eq:mlwf})-(\ref{eq:hr}) indicate lattice translations,
whereas for crystal structures with more than one atom per unit cell
$n$ and $m$ represent a combined orbital and site index, specifying
the various orbitals at all sites within the primitive unit cell. 

\subsection{Computational details}\label{ss:comp-details}

All results presented in this work are obtained from spin-polarized
first principles DFT calculations using the Quantum-ESPRESSO program
package,~\cite{quantum-espresso} the GGA exchange-correlation
functional of Perdew, Burke, and Ernzerhof,\cite{1996_perdew} and
Vanderbilt ultrasoft pseudopotentials.~\cite{1990_vanderbilt}
\mbox{La~(5$s$,5$p$)} and \mbox{Mn~(3$s$,3$p$)} semicore states are
included in the valence. The Hubbard $+U$ correction is applied using
the simplified approach according to Dudarev {\sl et
  al.},~\cite{1998_dudarev} which corresponds to the case $J=0$ in the
more elaborate expression by Lichtenstein {\sl et
  al.}.~\cite{Liechtenstein/Anisimov/Zaanen:1995} Projections on
orthogonalized atomic Mn($d$) orbitals are used to evaluate the $U$
dependent contributions to potential and energy.

To analyze the effect of the JT distortion on the electronic structure
of LaMnO$_3$, we perform GGA calculations with different degrees of
distortion. Starting from the ideal cubic perovskite structure, we
gradually increase the amplitude of the JT distortion
$|Q^x_\mathbf{R}|$ from 0 to $Q^{x}_{0}=0.151$~\AA. The latter value
corresponds to the amount of distortion found in the experimentally
observed crystal structure of LaMnO$_3$.\cite{1995_norby} For the
undistorted case we use a cubic perovskite structure with lattice
constant $a_0=3.9345$~\AA, which results in the same volume
\mbox{$V=60.91$~\AA$^3$} per formula unit as in the experimentally
observed $Pbnm$ structure.\cite{1995_norby} 

After obtaining the DFT Bloch bands, we construct MLWFs using the
\texttt{wannier90} program integrated into the Quantum-ESPRESSO
package.~\cite{Mostofi_et_al:2008} Starting from an initial projection
of atomic $d$ and $p$ basis functions centered on the Mn and O sites
onto the Bloch bands within an appropriately chosen energy window, we
obtain sets of either 14 atomic-like ($d$-$p$ model) or 2 more
extended $e_g$-like (effective $e_g$ model) MLWFs per spin channel and
unit cell. In both cases we will in the following refer to the MLWF
Hamiltonian matrix elements $h_{nm}^{\mathbf{0}}$ between the two
different $e_g$-like MLWFs located on the same Mn site as
\emph{on-site off-diagonal element} $q$, to any matrix element between
the same orbital at a particular site as \emph{on-site energy}
$\varepsilon$, and to matrix elements connecting two different sites
as \emph{hopping amplitudes} $t$. To assess the effect of the ``+$U$''
correction on the two different MLWF parameterizations we then
construct the corresponding MLWFs for the fully JT distorted structure
from GGA+$U$ calculations with different values for the Hubbard $U$.

Convergence of the DFT total energy and total magnetization has been
tested for the ideal cubic perovskite structure and ferromagnetic (FM)
order. We find the total energy converged to an accuracy better than
\mbox{1~mRy} and the total magnetization converged to an accuracy of
\mbox{0.05~$\mu_{\mathrm{B}}$} for a plane-wave energy cut-off of
\mbox{35~Ry} and a $\Gamma$-centered
\mbox{${10}\!\times\!{10}\!\times\!{10}$} k-point grid using a
Gaussian broadening of \mbox{0.01~Ry}. These values for plane-wave
cutoff and Gaussian broadening are used throughout this work, whereas
an appropriately reduced k-point grid of
\mbox{${7}\!\times\!{7}\!\times\!{10}$} is used for the JT distorted
structure with doubled unit cell in the $x$-$y$ plane. The MLWFs are
considered to be converged if the fractional change of the quadratic
spread (both gauge-invariant and non-gauge-invariant part) between two
successive iterations is smaller than $10^{-10}$.

\section{Results and Discussion}\label{s:resdis}

\subsection{$d$-$p$ TB parameterization for cubic
  LaMnO$_3$}\label{ss:cubic}

\begin{figure}
  \includegraphics[width=\columnwidth]{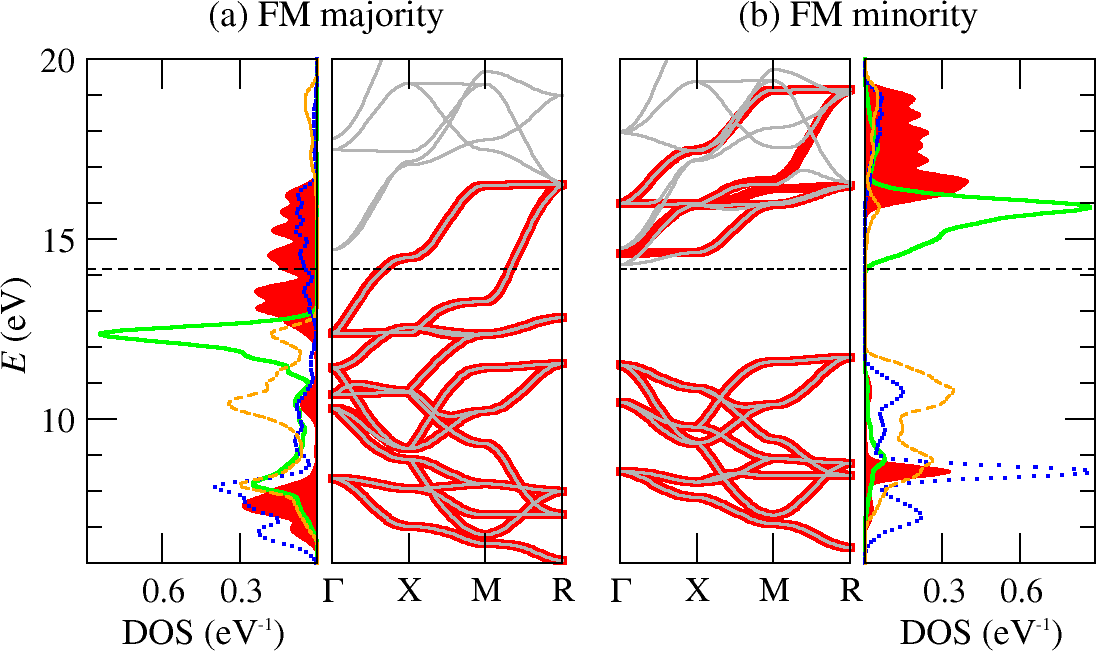}
  \caption{(Color online) Projected DOS and band structure along high
    symmetry lines within the BZ for cubic FM LaMnO$_3$. In the
    projected DOS plots, the filled (red) areas and solid (green)
    lines correspond to Mn($e_{g}$) and Mn($t_{2g}$) states,
    respectively, while dotted (blue) and dashed (orange) lines
    correspond to the O($p_\sigma$) and O($p_\pi$) states,
    respectively. In the band structure plots, the dispersion
    calculated from the MLWFs is represented by thick (red) lines.
    The Fermi level is indicated by the horizontal dashed lines.}
  \label{fig:c-pdos-bs}
\end{figure}

In this section we establish the general features of the extended
$d$-$p$ TB description of LaMnO$_3$ within the ideal cubic perovskite
structure, before we analyze the effect of the JT distortion in the
next section. We are considering a FM arrangement of magnetic moments,
but we have verified that the corresponding results for A-type
antiferromagnetic (A-AFM) order do not exhibit any significant
differences.

Fig.~\ref{fig:c-pdos-bs} shows the projected densities of states (DOS)
and band dispersion for both majority and minority spin channels. In
agreement with previous calculations it can be seen that the majority
spin DOS around the Fermi energy exhibit predominant Mn($e_{g}$)
orbital character, with Mn($t_{2g}$) and O($p$) bands located at
slightly lower
energies.\cite{Pickett/Singh:1996,Satpathy/Popovic/Vukajlovic:1996,Ederer/Lin/Millis:2007,Kovacik/Ederer:2010}
For minority spin, states with predominant Mn($e_{g}$) and
Mn($t_{2g}$) character are located above the Fermi level. Strong
hybridization between O($p$) and Mn($d$) orbitals is apparent from the
various peaks in the DOS around 8~eV.

\begin{figure}
  \includegraphics[width=0.9\columnwidth]{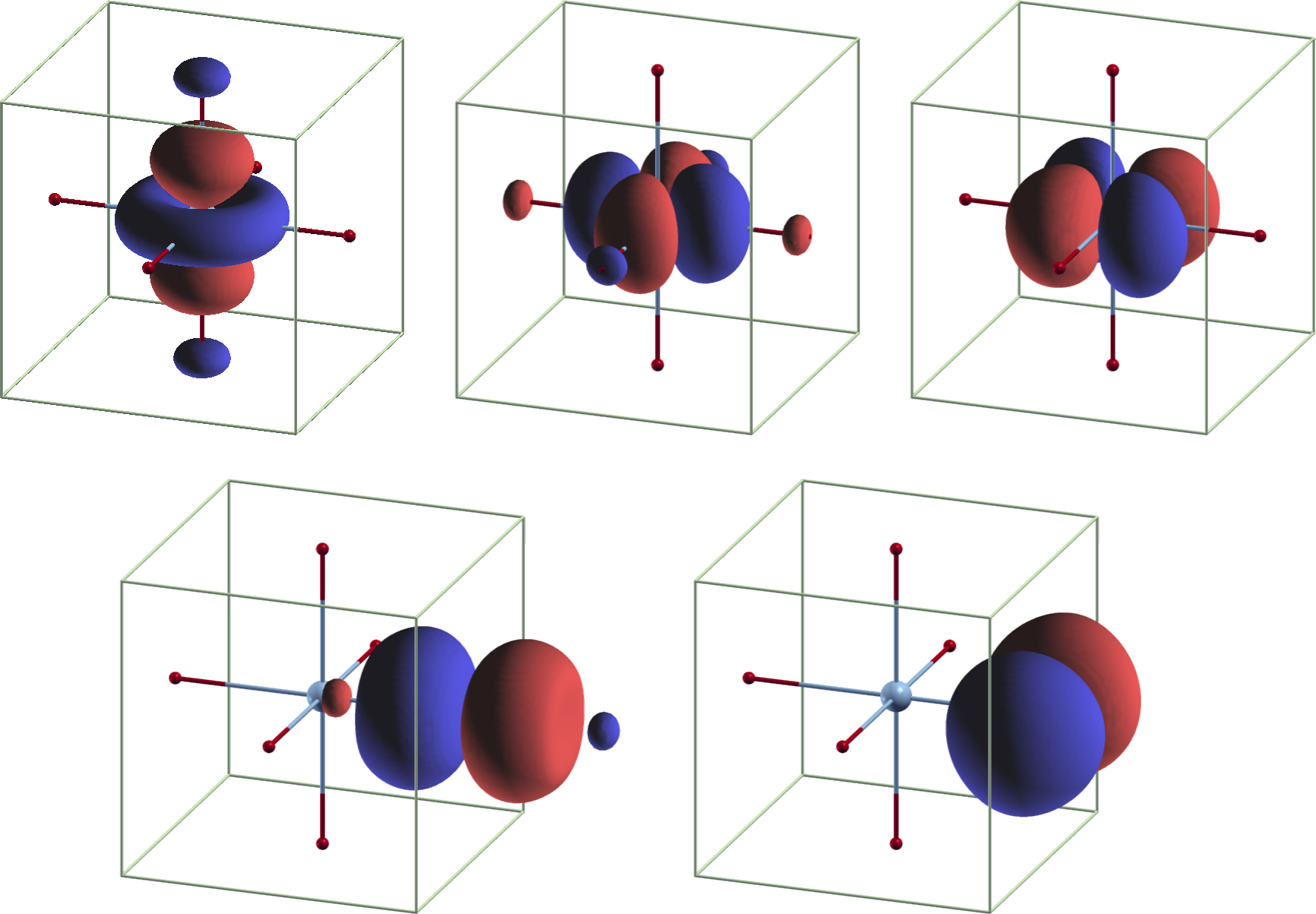}
  \caption{(Color online) Real space representation of the MLWFs
    corresponding to the $d$-$p$ parameterization (FM, majority
    spin). Depicted are Mn($3z^2-r^2$), Mn($x^2-y^2$), and Mn($xy$)
    MLWFs (top row from left to right) as well as two examples of
    O($p_{\sigma}$) and O($p_{\pi}$) orbitals (bottom row). The
    isosurface shown corresponds to a value of $\pm 1/\sqrt{V}$ where
    $V$ is the unit cell volume. Picture generated using
    XCRYSDEN.\cite{2003_kokalj}}
  \label{fig:dp-mlwf}
\end{figure}

We construct 14 MLWFs per spin channel from the Kohn-Sham states
located within an energy window of 3-17~eV and 3-20~eV for majority
and minority spin, respectively. This corresponds to a TB
representation of LaMnO$_3$ containing 5 Mn($d$) orbitals and 9 O($p$)
orbitals per unit cell (three $p$ orbitals corresponding to each of
the three oxygen atoms within the simple cubic perovskite unit
cell). The resulting MLWFs are depicted in
Fig.~\ref{fig:dp-mlwf}. Three of the O($p$) orbitals (one on each O
atom) are pointing towards the Mn atom and hybridize with the
Mn($e_{g}$) orbitals. We call them O($p_{\sigma}$) orbitals. The
remaining 6 O($p$) orbitals (two on each O atom) are oriented
perpendicular to the Mn-O bond and hybridize with the Mn($t_{2g}$)
orbitals. These will be called O($p_{\pi}$) orbitals in the following.

The MLWF bands (shown as thick (red) lines in
Fig.~\ref{fig:c-pdos-bs}) are identical to the corresponding DFT bands
for majority spin, whereas for minority spin the MLWF and DFT bands
above the Fermi energy exhibit certain differences which are due to
the strong entanglement with other bands in that energy region.

From the real-space Hamiltonian matrix elements in the MLWF basis,
Eq.~(\ref{eq:hr}), we find that the most dominant hopping corresponds
to the shortest Mn-O and O-O bonds (see Fig.~\ref{fig:1nhop}), but
that a variety of other hopping amplitudes are also non-negligible. In
the following we will focus on the nearest neighbor Mn-O hopping
amplitudes ($t_{eg,p\sigma}$ and $t_{t2g,p\pi}$) and analyze how they
are affected by the JT distortion and the inclusion of a Hubbard $U$
in the DFT calculations. A more detailed discussion of the $d$-$p$ TB
parameterization for cubic LaMnO$_3$ can be found in
Appendix~\ref{sec:appendix}.

\subsection{Effect of JT distortion on the $d$-$p$ TB
  parameterization}\label{ss:dp-jt}

We now analyze the effect of the JT distortion on the on-site energies
and nearest neighbor hopping parameters of the $d$-$p$ model. As in
the previous section, we will discuss only results for FM order as we
do not find significant differences for the case of A-AFM order.

\begin{figure}
  \centering
  \includegraphics[width=\columnwidth]{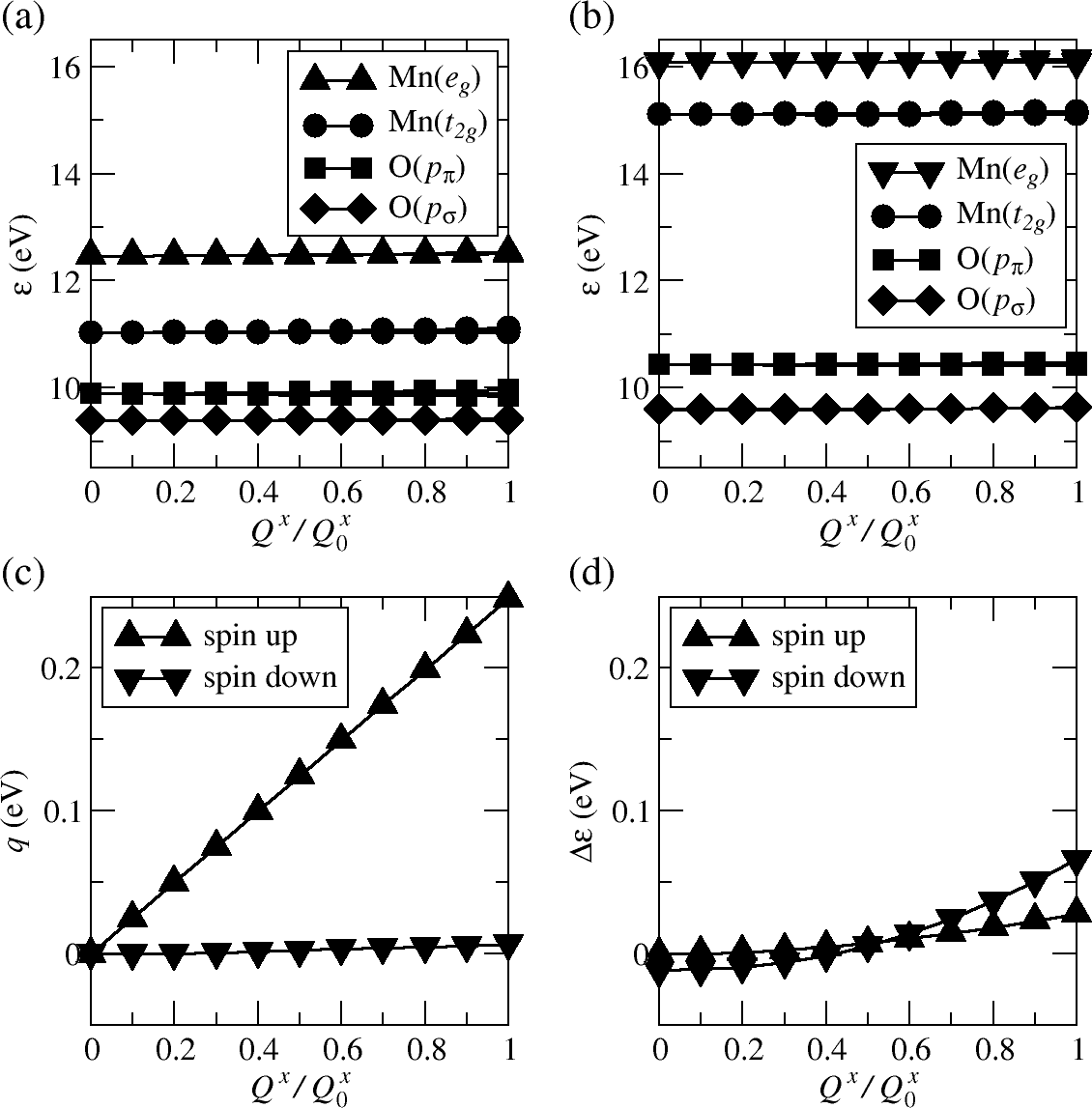}
  \caption{On-site MLWF matrix elements of the $d$-$p$ model as
    function of the JT distortion. (a)/(b): On-site energies for
    majority/minority spin. (c): On-site off-diagonal element
    $q$. (d): Splitting $\Delta\varepsilon$ between on-site energies
    of the two $e_g$-like MLWFs.}
  \label{fig:dp-jt-e}
\end{figure}

Fig.~\ref{fig:dp-jt-e} shows the diagonal on-site energies
$\varepsilon$ corresponding to Mn($e_g$), Mn($t_{2g}$), O(p$_\pi$),
and O($p_\sigma$) MLWFs, as well as the on-site off-diagonal matrix
elements $q$ between the two Mn($e_g$) orbitals on the same site as
function of the JT amplitude $Q^x/Q^x_0$. It can be seen that the
on-site energies for the various orbitals are essentially unaffected
by the JT distortion. The corresponding changes, which lead to small
differences between formerly symmetry-equivalent orbitals (see below
for the case of the two $e_g$-like MLWFs), are of the order of a few
tens of meV, and are thus negligible compared to the $e_g$-$t_{2g}$
splitting or the $d$-$p$ energy separation.

A more distinct effect of the JT distortion can be seen in the on-site
off-diagonal matrix element $q$ between the two majority spin
Mn($e_g$) orbitals on the same site. This matrix element is zero for
the undistorted structure, but exhibits a linear increase to a value
of about 0.25~eV for full JT distortion.  This is consistent with the
usual crystal-field picture of $e_g$ orbitals in an octahedral
environment (see also Eq.~(\ref{eq:jt})):\cite{Kanamori:1960} within a
$\{ |3z^2-r^2\rangle, |x^2-y^2\rangle \}$ orbital basis, the
$Q^x$-type JT distortion gives rise to a nonzero off-diagonal matrix
element that increases linearly with the JT distortion, while the
diagonal elements of the Hamiltonian remain constant.

However, the effect of the JT distortion observed for the minority
spin $q$ is significantly weaker than for the majority spin case. In
fact, it is of similar magnitude as the small splitting between the
corresponding diagonal matrix elements (on-site energies)
\mbox{$\Delta\varepsilon = \varepsilon[\text{Mn}(3z^2-r^2)] -
  \varepsilon[\text{Mn}(x^2-y^2)]$}.  We thus define a total
JT-induced orbital splitting within the ``$e_g$'' orbital subspace as
the difference in eigenvalues of the corresponding 2$\times$2 on-site
Hamiltonian matrix, which is given by
\mbox{$\delta=\sqrt{\Delta\varepsilon^2+4q^2}$}. From the results
presented in Fig.~\ref{fig:dp-jt-e} for full JT distortion,
i.e. $Q^x/Q^x_0=1$, we obtain $\delta_\uparrow=0.50$~eV for majority
spin and $\delta_\downarrow=0.07$~eV for minority spin.

To understand this large difference between majority and minority
spin, we note that in a partially covalent system like LaMnO$_3$, the
JT-induced orbital splitting is caused by a superposition of two
effects: i) the true (electro-static) crystal-field effect, and ii) a
``ligand-field'' effect due to changes in hybridization with the
surrounding ligand orbitals. While the pure electro-static
crystal-field effect is identical for both spin projections, the
hybridization with the surrounding ligand orbitals is different for
majority and minority spin orbitals, due to their large energy
difference of about 3-4~eV.

In order to verify whether the large difference between majority and
minority spin JT-splitting is indeed due to a strong ligand-field
contribution, we have constructed an alternative set of 20 MLWFs per
formula unit and spin channel, where we explicitly included also the
bands corresponding to O($s$) and semi-core La(5$p$) states. These
states are energetically lower than the Mn($d$) and O($p$) bands shown
in Fig.~\ref{fig:c-pdos-bs}, and are centered around $E \approx -3$~eV
(La(5$p$)) and $E \approx -5$~eV (O($s$)). The inclusion of these
states in the MLWF basis reduces the ``tails'' in the $e_g$-like MLWFs
located at surrounding O and La atoms, and should therefore decrease
the ligand-field contribution to the local JT splitting. Indeed, we
obtain $\delta_{\uparrow/\downarrow} = 0.36/0.12$~eV in this case,
i.e. the spin-dependent ligand-field contribution to the JT splitting
is indeed significantly reduced compared to the $d$-$p$ MLWFs. The
remaining difference can be ascribed to further contributions of
orbitals that are not included in the MLWF basis and to the remaining
small O($s$), O($p$), and La($p$) contributions on the surrounding
sites, which are required to ensure orthogonality between the MLWFs.

We also note that the JT splitting corresponding to the rather
localized $e_g$ MLWFs of the $d$-$p$ TB model (0.50~eV for majority
spin) is a factor of two smaller than the corresponding splitting for
the more spatially extended MLWFs of the effective $e_g$ model which
was reported in Ref.~\onlinecite{Kovacik/Ederer:2010} (0.97~eV for
majority spin). This demonstrates the much stronger ligand-field
effect in the less localized effective $e_g$ MLWFs.

\begin{figure}
  \centering
  \includegraphics[width=\columnwidth]{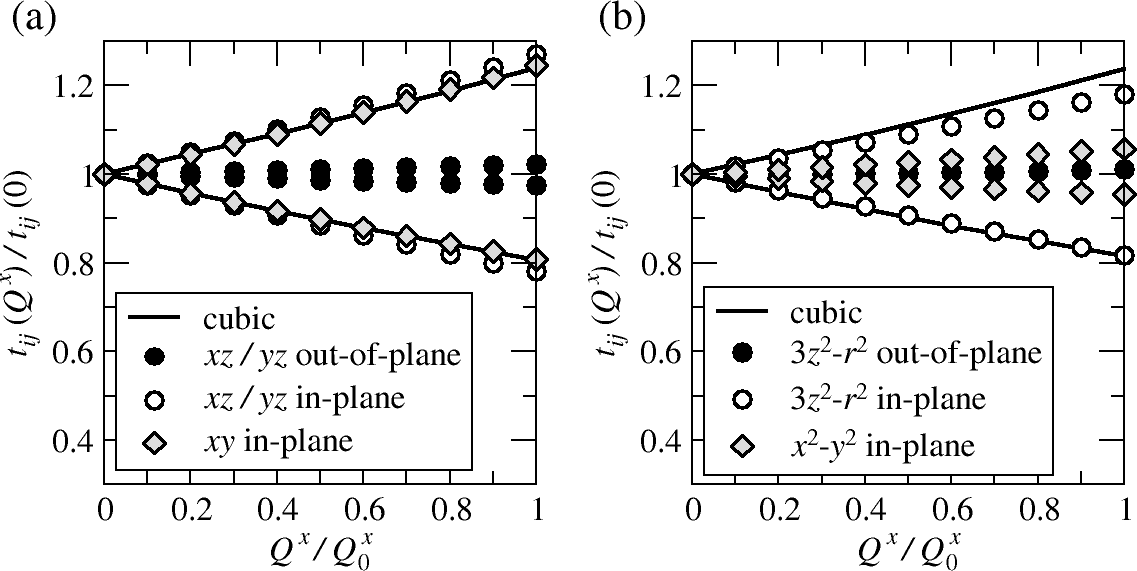}
  \caption{Effect of increasing JT distortion on the hopping
    amplitudes $t_{t2g,p\pi}$ (a) and $t_{eg,p\sigma}$ (b) between
    Mn($d$) and O($p$) nearest-neighbor orbitals (symbols). The
    different hopping amplitudes are denoted via the corresponding
    Mn($d$) orbital and the direction of the hopping (``in-plane''
    corresponds to $x$-$y$ directions, ``out-of-plane'' to the
    $z$-direction). The solid lines correspond to the hopping
    amplitudes for a perfect cubic perovskite structure with lattice
    constant $a = a_0 \pm \Delta a$, where $\Delta a = \sqrt{2}
    Q^x$. Upper/lower branches correspond to short/long Mn-O bonds.}
  \label{fig:dp-jt-tupdprel}
\end{figure}

Next, we analyze the effect of the JT distortion on the nearest
neighbor hopping between Mn($d$) and O($p$)
orbitals. Fig.~\ref{fig:dp-jt-tupdprel} shows the corresponding
changes in the majority spin hopping amplitudes (the results for
minority spin do not exhibit any qualitative differences).

We first discuss hopping between Mn($t_{2g}$) and O($p_\pi$) orbitals
(Fig.~\ref{fig:dp-jt-tupdprel}a). It can be seen that the hopping
amplitude $t_{t2g,p\pi}$ splits essentially according to the
different Mn-O bond lengths in the JT distorted structure. The hopping
amplitudes along the short (long) Mn-O bonds within the $x$-$y$ plane
increase (decrease) with increasing JT distortion, while the hopping
along the $z$-direction, i.e. corresponding to constant Mn-O bond
distance, is only weakly affected.

In order to assess to what extent the $d$-$p$ hoppings in the JT
distorted structure are simply determined by the corresponding Mn-O
distances, we also calculate MLWFs for undistorted cubic LaMnO$_3$
with different lattice constants $a$. The Mn-O distances in the cubic
structures with $a=a_0 \pm \Delta a$ are identical to the long/short
Mn-O bond length in the JT distorted structure with $Q^x =
\Delta{a}/\sqrt{2}$. The corresponding results for the nearest
neighbor hopping amplitudes between Mn($d$) and O($p$) orbitals are
also shown in Fig.~\ref{fig:dp-jt-tupdprel}.

It is apparent that the $t_{t2g,p\pi}$-type hopping amplitudes in the
JT distorted structures are nearly identical to the corresponding
hopping amplitude in the undistorted cubic structure with the same
Mn-O distance. The small deviations between these two cases as well as
the weak effect of the JT distortion on the $t_{t2g,p\pi}$ hopping
along $z$ are due to changes in the orbital character of the MLWFs
with increasing JT distortion. Thus, the magnitudes of the various
hopping amplitudes are indeed determined mostly by the corresponding
Mn-O bond lengths.

The case of the $t_{eg,p\sigma}$ hopping is only slightly different.
The hopping (both in-plane and along $z$) between the
Mn($3z^2-r^2$)-type orbital and the surrounding O($p_\sigma$) orbitals
compares well with the hopping amplitude in the undistorted structure
with the same Mn-O distance (even though the agreement for the short
Mn-O distance is not as good as for $t_{t2g,p\pi}$). On the other
hand, the JT-induced splitting of the in-plane hopping between the
Mn($x^2-y^2$)-type orbital and the surrounding O($p$) orbitals is
significantly weaker than the corresponding bond-length dependence in
the cubic structure. This indicates a strong change of the
$\vert{x^2-y^2}\rangle$-type orbital with increasing JT
distortion. This change is due to different admixture of other
orbitals that are not explicitly included in the MLWF construction,
i.e. O($s$) and La($p$), which lead to a reduction/expansion of the
lobes directed along the shorter/longer Mn-O bonds with increasing JT
distortion. This partially compensates the effect of changing Mn-O
distance and leads to the observed JT dependence of the hopping.

\subsection{Effect of $U$ within the $d$-$p$ model}\label{ss:dp-jt-u}

\begin{figure}
  \centering
  \includegraphics[width=\columnwidth]{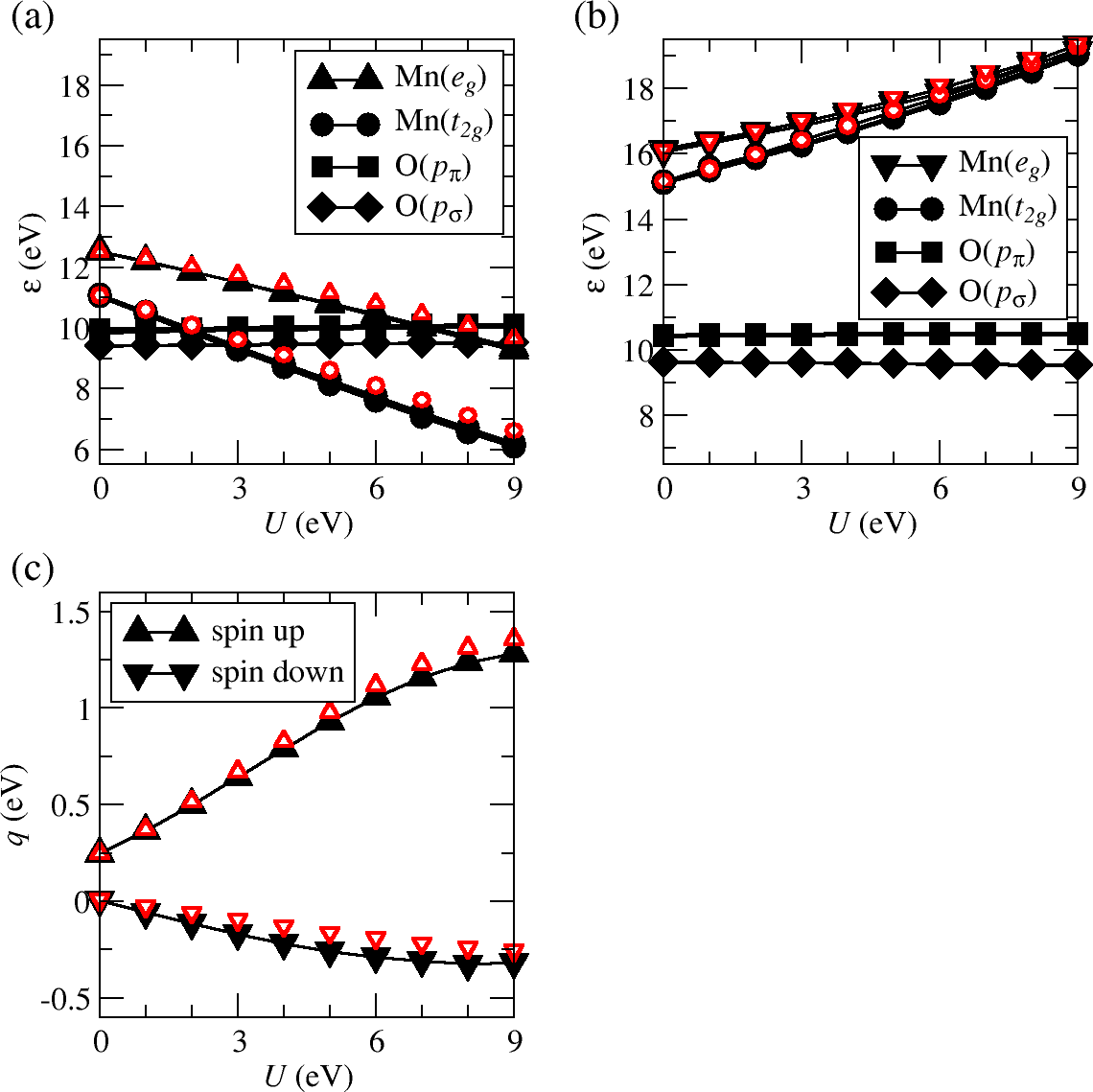}
  \caption{(Color online) On-site energies corresponding to the
    $d$-$p$ model as a function of Hubbard $U$ for majority spin (a)
    and minority spin (b). (c): $U$ dependence of the on-site
    off-diagonal elements $q$. The small open (red) symbols represent
    the GGA+$U$ potential shifts added to the corresponding MLWF
    on-site matrix elements for $U=0$.}
  \label{fig:dp-jt-ediagu}
\end{figure}

In order to investigate the influence of the Hubbard $U$ parameter on
the TB parameterization obtained from the MLWFs, we now perform
GGA+$U$ calculations for the fully JT distorted structure and FM order
using different values for $U$. Figure~\ref{fig:dp-jt-ediagu} shows
the resulting $U$ dependence of the on-site MLWF matrix elements. It
can be seen that the diagonal elements (on-site energies
$\varepsilon$) corresponding to O($p$)-type MLWFs are virtually
unaffected by the value of $U$, while the corresponding matrix
elements for Mn($t_{2g}$) and Mn($e_{g}$) orbitals exhibit a more or
less linear dependence on $U$.

This linear dependence is a direct consequence of the $U$ dependent
potential shift applied to the TM $d$ states within the DFT ``+$U$''
approach:\cite{1998_dudarev}
\begin{equation}
\label{eq:ldau}
\Delta V^\sigma_{mm'} = U (\frac{\delta_{mm'}}{2}-n^\sigma_{mm'}) \quad .
\end{equation}
Here, $n^\sigma_{mm'}$ is the occupation matrix element between atomic
orbitals $m$ and $m'$ for spin projection $\sigma$, which is
calculated from the projection of the occupied Bloch states on fixed
atomic orbitals. 

In order to quantify, whether the $U$ dependence of the on-site MLWF
matrix elements does indeed correspond to this potential shift, we
explicitly evaluate Eq.~(\ref{eq:ldau}) for each $U$, using the atomic
occupation matrix elements obtained from the corresponding GGA+$U$
calculation, and add the so-obtained shifts to the on-site MLWF matrix
elements for $U=0$~eV. The resulting data is shown as small open (red)
symbols in Fig.~\ref{fig:dp-jt-ediagu} and is nearly identical to the
MLWF matrix elements calculated for the corresponding values of $U$.
Thus, the GGA+$U$ potential shifts are directly reflected in the
on-site matrix elements of the MLWFs.

\begin{table}
\caption{Average occupations $n_{mm}^\sigma$ corresponding to
  Mn($t_{2g}$) and Mn($e_g$) orbitals calculated for $U=0$ in the
  fully JT distorted structure. Rows denoted ``atomic'' correspond to
  the fixed atomic orbitals used to evaluate the GGA+$U$ functional;
  rows denoted ``MLWF'' contain the occupation of the MLWFs calculated
  according to Eq.~(\ref{eq:n-mlwf}); rows denoted ``formal''
  correspond to the ionic limit based on a high-spin $d^4$
  configuration of the Mn cation. The last column contains the
  off-diagonal occupation matrix element $n_{mm'}^\sigma$ between the
  two different $e_g$ orbitals. The first/last three rows correspond
  to majority/minority spin.}
  \label{tab:occupations}
\begin{ruledtabular}
\begin{tabular}{lcccc}
 & $\sigma$ & $t_{2g}$ & $e_g$ & $e_g$ off diagonal \\
\hline
atomic & $\uparrow$ & 0.99 & 0.72 & $-$0.11 \\
MLWF   & $\uparrow$ & 1.00 & 0.68 & $-$0.13 \\
formal & $\uparrow$ & 1.0 & 0.5 & --- \\
\hline
atomic & $\downarrow$ & 0.11 & 0.22 & 0.03 \\
MLWF   & $\downarrow$ & 0.11 & 0.15 & 0.02 \\
formal & $\downarrow$ & 0.0 & 0.0 & ---\\
\end{tabular}
\end{ruledtabular}
\end{table}

The reason for this good correspondence between the GGA+$U$ potential
shifts and the $U$ dependence of the on-site MLWF matrix elements is
the fact that the MLWFs of the extended $d$-$p$ model are rather
similar to the atomic orbitals used as projector functions within the
GGA+$U$ approach. To further demonstrate this similarity, we also
compare the occupation matrix elements used to evaluate the GGA+$U$
potential shifts with the occupation of the corresponding
MLWFs:\cite{footnote:spin-index}
\begin{equation}
  \label{eq:n-mlwf}
  n^\text{MLWF}_{mm'} = \int_{-\infty}^{E_{\text{F}}} \mathrm{d}\epsilon \int_\text{BZ} \mathrm{d}\mathbf{k}
  \sum_{l} \left(U^{(\mathbf{k})}_{lm}\right)^* \delta(\epsilon-\epsilon_{l\mathbf{k}}) \, U^{(\mathbf{k})}_{lm'}
  \,,
\end{equation}
where $E_{\text{F}}$ is the Fermi energy. The corresponding values for
$U=0$~eV are listed in Table~\ref{tab:occupations}. It is apparent
that the occupations of the MLWFs are very similar to the occupations
of the atomic orbitals used as GGA+$U$ projector
functions. Furthermore, we note that due to the strong hybridization
between the Mn($d$) and O($p$) states, the occupation of the
atomic-like $d$ orbitals are quite different from a na\"ive
expectation based on the formal ionic configuration of the Mn$^{3+}$
cation.

The $U$ dependence of the on-site off-diagonal matrix element $q$ can
be understood in the same way and is related to the off-diagonal
occupation matrix element $n_{mm'}^\sigma$, where $m\, \widehat{=}\,
|3z^2-r^2\rangle$ and $m'\, \widehat{=}\, |x^2-y^2\rangle$. The
nonlinearities that can be observed for large $U$ are due to changes
in the corresponding $n_{mm'}^\sigma$. Note that since $q$ is directly
related to the JT splitting, increasing $U$ effectively amounts to
increasing the strength of the JT coupling.

\begin{figure}
  \centering
  \includegraphics[width=\columnwidth]{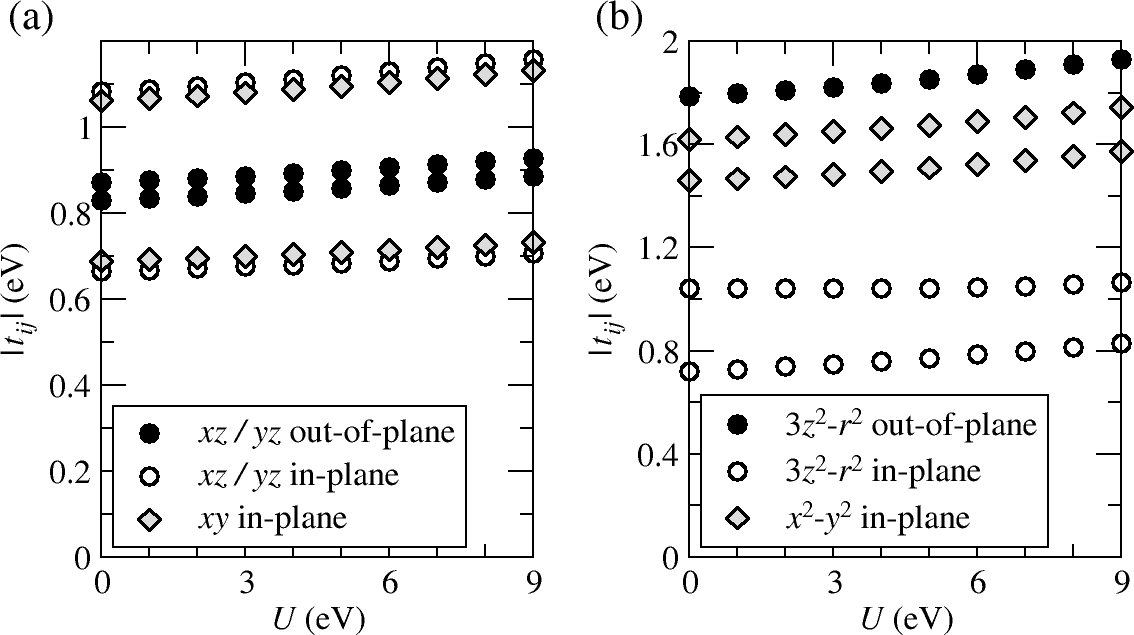}
  \caption{Magnitude of hopping amplitudes between Mn($d$) and O($p$)
    nearest-neighbor orbitals as function of the Hubbard $U$. (a):
    $t_{2g}-p_{\pi}$ hoppings, (b): $e_g-p_{\sigma}$ hoppings. Notation
    is the same as in Fig.~\ref{fig:dp-jt-tupdprel}.}
  \label{fig:dp-jt-tupdprelu}
\end{figure}

The $U$ dependence of the hopping amplitudes between O($p$) and
Mn($d$)-like MLWFs for majority spin and FM order is shown in
Fig.~\ref{fig:dp-jt-tupdprelu}. All depicted hopping amplitudes
exhibit a similar small increase in magnitude. The effect of $U$ on
the hopping amplitudes for minority spin (not shown) is even weaker
than for the majority spin orbitals. These small changes in the
hopping amplitudes for both majority and minority spin can be
attributed to changes in orbital character of the corresponding
MLWFs. These changes in orbital character arise from a different
admixture (hybridization) of orbitals centered at the surrounding
ions, which results from the $U$ dependent shifts of the on-site
energies of the Mn($d$) states relative to all other orbitals.

\subsection{Effect of $U$ within the effective $e_g$ model}\label{ss:dd-jt-u}

We now contrast the $U$ dependence of the TB parameters for the
extended $d$-$p$ model presented in the previous section with the case
of an effective $e_g$ model with only two $e_g$-like Wannier orbitals
on each Mn site. This is probably the most common model used to
describe the physics of manganites (see
e.g. Ref.~\onlinecite{Dagotto/Hotta/Moreo:2001}), and a detailed
analysis of the effect of various distortions and different magnetic
arrangements on the corresponding MLWF matrix elements for $U=0$~eV
has been presented in Ref.~\onlinecite{Kovacik/Ederer:2010}. Here we
focus on changes in the MLWF parameterization due to the Hubbard $U$
correction for the purely JT distorted structure and FM order. The
corresponding Wannier functions are constructed from the Kohn-Sham
states located within an energy window of 12.0-17.0~eV and
15.9-20.0~eV for majority and minority spin, respectively. For further
details see Ref.~\onlinecite{Kovacik/Ederer:2010}.

\begin{figure}
  \centering
  \includegraphics[width=\columnwidth]{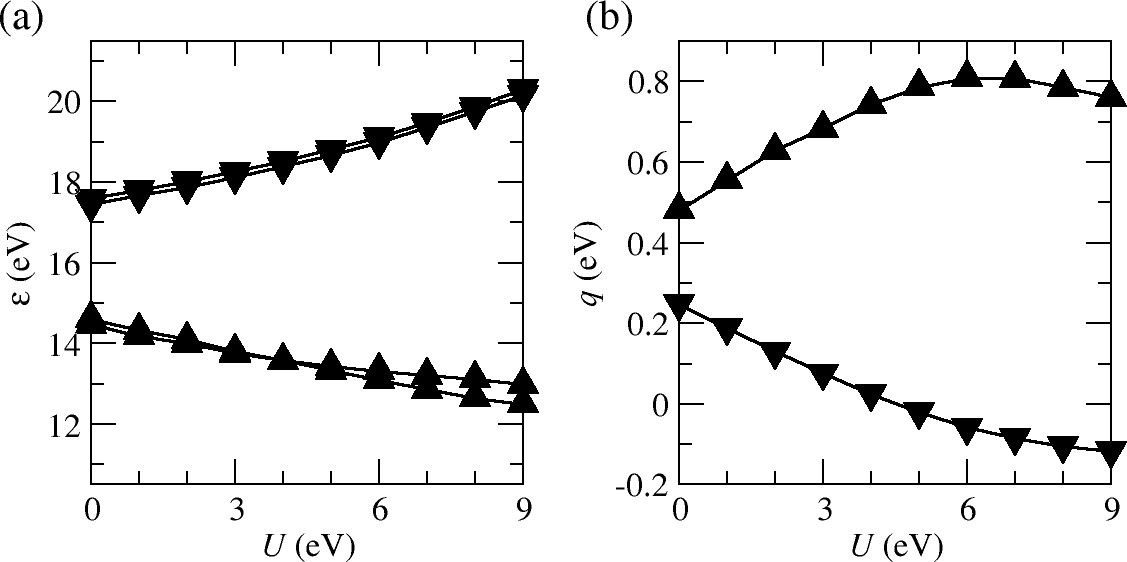}
  \caption{On-site energies $\varepsilon$ (a) and on-site off-diagonal
    matrix element $q$ (b) of the effective $e_g$ MLWF
    parameterization as function of the Hubbard $U$. Majority spin and
    minority spin is denoted by up and down triangles, respectively.}
  \label{fig:dd-jt-eu}
\end{figure}

Fig.~\ref{fig:dd-jt-eu} shows the $U$ dependence of the on-site MLWF
Hamiltonian matrix elements for the effective $e_g$ model. It can be
seen that in particular the on-site energies exhibit a very similar
trend as the on-site energies of the $e_g$-like MLWFs in the $d$-$p$
model, but that the $U$ dependence is weaker than in the latter
case. The slope $\text{d}\varepsilon/\text{d}U$ for both majority and
minority spin is only about 66\,\% compared to the $d$-$p$ model. This
indicates that the $U$ dependence of the on site energies for the
effective $e_g$ MLWFs is still determined by the GGA+$U$ potential
shifts, Eq.~(\ref{eq:ldau}), but is renormalized by the extent of
overlap between the extended MLWF and the corresponding $e_g$ atomic
orbital. In other words, the $U$ dependent energy shift experienced by
the MLWF is determined by the projection of the MLWF on atomic
orbitals and the occupation of these atomic orbitals.

In particular, this shows that the effect of $U$ on the on-site
energies of the effective $e_g$ MLWFs are \emph{not} determined by the
occupation of the MLWFs themselves via a relation similar to
Eq.~(\ref{eq:ldau}). The occupation of the effective $e_g$ MLWFs are
essentially identical to the ``formal'' occupations listed in
Table~\ref{tab:occupations}, corresponding to the formal $3+$ charge
state of the Mn cation in LaMnO$_3$ (0.5 for majority spin and 0.0 for
minority spin). Therefore, the occupation-dependent potential shift
following from a mean-field approximation to the Hubbard interaction
similar to Eq.~(\ref{eq:ldau}) would be zero for the majority spin
$e_g$ states.\cite{footnote:j0} The $U$ dependence of the
corresponding on-site energies is thus notably different from a
mean-field treatment of the electron-electron interaction in a
two-orbital Hubbard-like model derived from the effective $e_g$ TB
parameterization.

We point out that in addition to the different dependence on orbital
occupation, the magnitude of the screened Hubbard $U$ acting on the
extended $e_g$-like Wannier orbitals will of course be significantly
reduced compared to the magnitude of $U$ corresponding to more
localized atomic-type orbitals. A comparison of the value of $U$ for
``atomic-like'' and ``effective'' Wannier orbitals in an Fe-pnictide
system based on calculations using constrained random phase
approximation has been presented recently in
Ref.~\onlinecite{Aichhorn_et_al:2009}.

Similar to the case of the on-site energies, the $U$ dependence of the
off-diagonal on-site matrix elements $q$, shown in
Fig.~\ref{fig:dd-jt-eu}b, resembles the $U$ dependence of the
corresponding atomic-like MLWFs shown in
Fig.~\ref{fig:dp-jt-ediagu}c. The decrease of $q$ for majority spin
observed for $U>6$~eV can then be ascribed to changes in orbital
character of the extended MLWFs, which lead to different projections
between MLWFs and the corresponding atomic orbitals.

\begin{figure}
  \centering
  \includegraphics[width=\columnwidth]{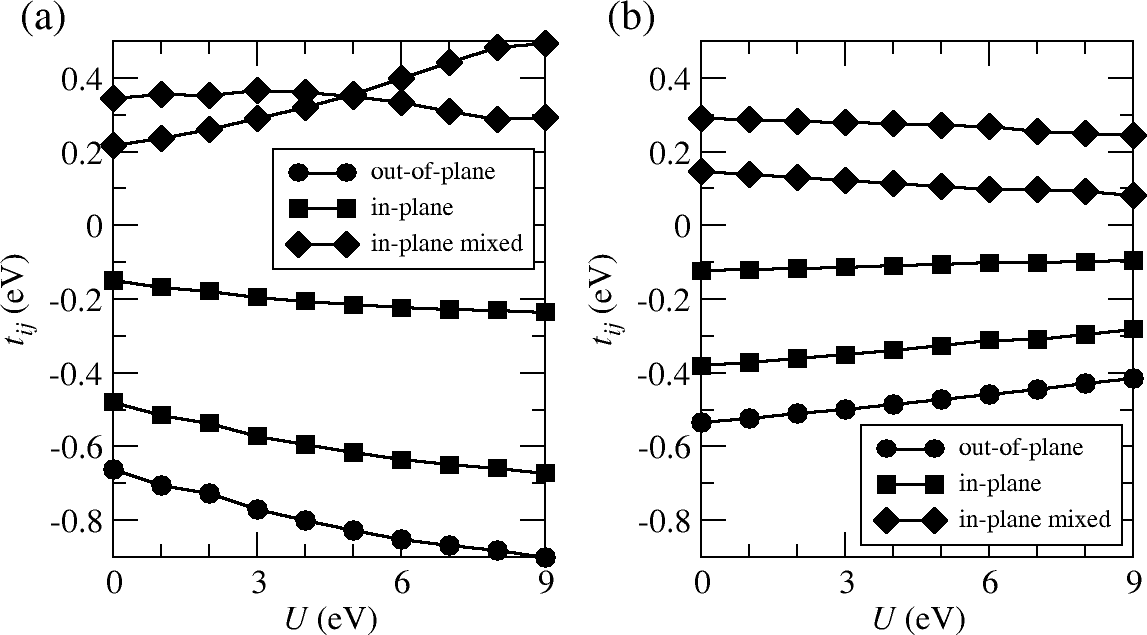}
  \caption{Effect of $U$ on the nearest neighbor hopping amplitudes of
    the effective $e_g$ MLWFs for majority spin (a) and minority spin
    (b). ``In-plane'' and ``out-of-plane'' corresponds to hopping
    between similar orbitals within the $x$-$y$ plane and along $z$,
    respectively, while ``in-plane mixed'' denotes the hopping between
    two different $e_g$ MLWFs at neighboring sites within the $x$-$y$
    plane.}
  \label{fig:dd-jt-tu}
\end{figure}

Finally, Fig.~\ref{fig:dd-jt-tu} shows the $U$ dependence of the
nearest neighbor hopping parameters for the effective $e_g$ MLWFs. For
majority spin, there is an overall increase in the magnitude of the
hopping, whereas for minority spin there is an overall
decrease. Compared to the $d$-$p$ model discussed in the previous
section, the effect of $U$ on the hopping amplitudes of the effective
$e_g$ model is strongly enhanced, with typical changes between 30\,\%
and 60\,\% over the considered range of $U$ values. For the in-plane
hopping between $|3z^2-r^2\rangle$ and $|x^2-y^2\rangle$-type orbitals
the change can be even more than 100\,\%.

This strong effect of $U$ on the hopping amplitudes of the effective
$e_g$ MLWFs can be explained by the changes in the underlying $d$-$p$
model. The effective $e_g$ MLWFs are essentially linear combinations
of an atomic-like Mn($e_g$) orbital and O($p_\sigma$) orbitals on the
surrounding anions. As demonstrated in the previous section, the
on-site energies of the atomic $d$ orbitals are strongly shifted
relative to the oxygen $p$ orbitals as a function of $U$. This shift
affects the hybridization between the atomic orbitals and thus changes
the admixture of O($p_\sigma$) orbitals in the effective $e_g$
MLWF. For majority spin, increasing $U$ decreases the energy
difference between atomic O($p$) and Mn($d$) orbitals and thus
increases the hybridization, i.e. the admixture of O($p$) in the
effective $e_g$ MLWF. This leads to an increase of the hopping
amplitude due to the larger overlap between effective $e_g$ orbitals
at neighboring Mn sites. For minority spin the trend is opposite to
this, resulting in the decrease in magnitude seen in
Fig.~\ref{fig:dd-jt-tu}b.

Considering a simple $d$-$p$ nearest neighbor TB model for the cubic
perovskite structure, one can show that in the limit of large energy
separation between the $d$ and $p$ orbitals, the ``antibonding'' bands
that result from the $d$-$p$ hybridization are formally equivalent to
an effective ``$d$-only'' TB model with direct hopping between
$d$-orbitals on adjacent TM sites. The amplitude of this ``effective''
$d$-$d$ hopping is thereby given as:
\begin{equation}
  t^\text{eff}_{dd} = \frac{t^2_{dp}}{\varepsilon_d-\varepsilon_p}
  \quad .
  \label{eq:teff}
\end{equation}
Obviously, the overall trends discussed in the previous paragraph are
consistent with this simple picture, but the limit for which
Eq.~(\ref{eq:teff}) is valid is not fulfilled in the case of
LaMnO$_3$. This is apparent from the on-site energies shown in
Fig.~\ref{fig:dp-jt-ediagu}a, where it can be seen that for values of
$U$ larger than 5-6~eV the Mn($e_g$) states become essentially
degenerate with the O($p$) states. We also verified that
Eq.~(\ref{eq:teff}) is not fulfilled quantitatively by the hoppings
calculated from both sets of MLWFs ($d$-$p$ and effective
$e_g$). Thus, even though it is possible to obtain a reasonable MLWF
parametrization of the effective $e_g$ bands for all values of $U$, it
is not obvious whether a low-energy description of LaMnO$_3$ based
only on effective $e_g$ states is always physically reasonable.

The partial ``breakdown'' of the effective $e_g$ description of
LaMnO$_3$ for values of $U$ larger than 5-6~eV also leads to a strong
dependence of some MLWF parameters on subtle differences in the
underlying Kohn-Sham bandstructure. For example the ``crossing'' of
the mixed in-plane hopping amplitudes, i.e. corresponding to the
in-plane hopping between $|3z^2-r^2\rangle$ and $|x^2-y^2\rangle$-type
orbitals, that can be seen around $U=5$~eV in
Fig.~\ref{fig:dd-jt-tu}a, appears only if orthogonalized atomic
orbital are used to evaluate the ``+$U$'' correction to the GGA energy
functional. If instead a projection on non-orthogonal atomic orbitals
is used (option ``atomic'' instead of ``ortho-atomic'' within
QuantumESPRESSO), the resulting MLWF matrix elements do not exhibit
this feature.

\section{Summary and Conclusions}\label{sec:summary}

In summary, we have discussed differences in the MLWF-derived TB
parameterization of LaMnO$_3$ that result from different values of the
Hubbard $U$ used in the GGA+$U$ calculation from which the MLWFs are
obtained. Thereby, we have compared two different ways to represent
the important bands around the Fermi energy. First, a \emph{$d$-$p$ TB
  model} based on atomic-like Mn($d$) and O($p$) MLWFs, and second, an
\emph{effective $e_g$ TB model} involving only two $e_g$-like Wannier
orbitals per Mn.

We have shown that the hopping amplitudes of the $d$-$p$ model are
only weakly affected by a variation of $U$, and that the resulting
changes in the on-site Hamiltonian matrix elements are consistent with
a mean-field approximation to the electron-electron interaction for
the corresponding TB Hubbard Hamiltonian. As a result, the TB
parameters for the $d$-$p$ model of LaMnO$_3$ are fairly insensitive
to variations of $U$ and can therefore be determined accurately
without detailed knowledge about the precise value of $U$ for the
corresponding $d$ orbitals.

In contrast, the hopping amplitudes for the effective $e_g$ TB
parameterization depend strongly on the value of $U$ used in the
GGA+$U$ calculation. This is due to pronounced changes in the amount
of admixture of O($p$) ligand orbital character in the effective $e_g$
MLWFs, which in turn is due to the large $U$ dependent shifts of the
on-site energies of the corresponding atomic orbitals. Furthermore,
the $U$ dependence within a mean-field Hubbard Hamiltonian derived
from the effective $e_g$ MLWF parameters is distinctly different from
the $U$ dependence of the MLWF parameters themselves. The former is
determined by the occupations of the effective $e_g$ orbitals and the
value of the screened Hubbard interaction in that basis, whereas the
latter is determined by the occupations and the value of the screened
Hubbard interaction corresponding to atomic-like $d$ states. A
determination of suitable TB parameters for an effective $e_g$ model
of LaMnO$_3$ therefore requires an accurate knowledge of $U$ in both
basis sets. This is particularly important for the determination of
the JT coupling strength, which is determined from the on-site
splitting within the $e_g$ orbital manifold. This splitting on the
other hand is critically affected by the value of $U$ in both the
GGA+$U$ calculation and the mean-field Hubbard Hamiltonian.

In addition to the $U$ dependence, we have also analyzed the effect of
the staggered JT distortion (``$Q^x$-type'') on the TB
parameterization of the $d$-$p$ model. We have found that the JT
distortion manifests itself both as a local ligand-/crystal-field
splitting within the $e_g$ orbital manifold, as well as via a
pronounced bond-length dependence of the nearest neighbor $d$-$p$
hopping amplitudes. Our results demonstrate that even for the rather
localized atomic-like $d$-$p$ MLWFs, the ligand-field effect dominates
over the purely electro-static crystal-field effect. Furthermore, we
have verified that the changes in the nearest neighbor hopping
amplitudes due to the JT distortion are almost fully determined by the
resulting change in Mn-O bond length.

A number of more general conclusions can be drawn from the results
obtained in this work. It is reasonable to assume that the
sensitivity/insensitivity of certain MLWF parameters from the specific
value of $U$ used in the GGA+$U$ calculation can be generalized to a
general sensitivity/insensitivity from the choice of
exchange-correlation functional used in the electronic structure
calculation. In the present study, we made use of the similarity
between the ``+$U$'' correction to the GGA functional and the
mean-field approximation to the electron-electron interaction in the
corresponding model Hamiltonian (assuming of course that the latter is
described via a local Hubbard interaction). Using this similarity, the
observed trends in the MLWF TB parameters, and the difference between
the two different TB models can be explained. The same similarity
between DFT and mean-field Hubbard model is not present if the
electronic structure is calculated using other ``beyond LDA/GGA''
methods such as for example self-interaction corrected
methods,\cite{Perdew/Zunger:1981} hybrid functionals,\cite{Becke:1993}
or the GW approach.\cite{Aryasetiawan/Gunnarsson:1998} In these cases
the electronic band-structure should also be compared with the
mean-field approximation of an appropriate TB Hamiltonian. To extract
the ``bare'' or ``non-interacting'' on-site energies and off-diagonal
on-site matrix elements, one has to then subtract a $U$ dependent
shift similar to Eq.~\ref{eq:ldau}. The corresponding value of $U$ has
to be calculated by other means, for example via the recently proposed
approach based on constrained random phase
approximation.~\cite{Aryasetiawan_et_al:2004,Miyake/Aryasetiawan:2008}

While the so-obtained TB models will always give an essentially
perfect representation of the Kohn-Sham bands for the given reference,
it is still important to test whether the corresponding
parameterization is also transferable to slightly different
configurations, i.e. with different structural distortions and/or
magnetic/orbital order. Of course, as shown through the comparison
between the effective $e_g$ and the more elaborate $d$-$p$ model for
LaMnO$_3$, transferability can always be improved by including more
orbitals in the TB basis set.

Finally, we note that in the present work we have focused on how to
obtain TB parameters using the DFT Kohn-Sham band-structure as
(mean-field) reference. An entirely different, albeit at least equally
important, question is whether these bands, calculated using a
suitable exchange-correlation functional for a particular system, are
indeed a good representation of the real material. However, this
question is beyond the scope of the present study and will most likely
be the topic of future research for a number of years to come.

\begin{acknowledgments}
  This work was supported by Science Foundation Ireland under
  Ref.~SFI-07/YI2/I1051 and made use of computational facilities
  provided by the Trinity Center for High Performance Computing. We
  also acknowledge partial support by the EU-FP7 project ATHENA.
\end{acknowledgments}

\appendix

\section{Further details on the $d$-$p$ TB parameterization for cubic LaMnO$_3$} \label{sec:appendix} 

\begin{figure}
  \includegraphics[width=0.75\columnwidth]{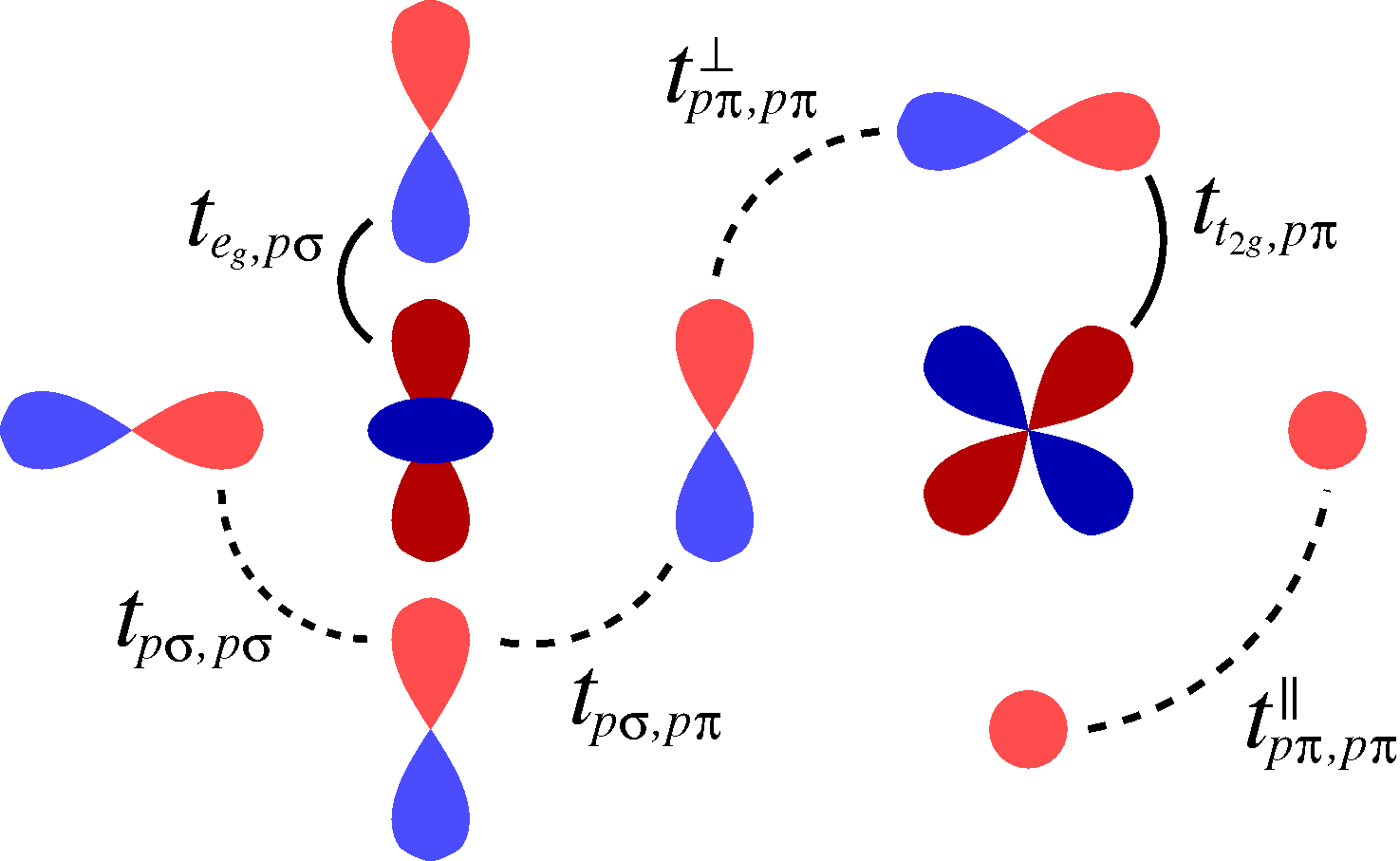}
  \caption{(Color online) Schematic depiction of all closest neighbor
    Mn-O and O-O hopping amplitudes within the extended $d$-$p$ model
    of LaMnO$_3$.}
  \label{fig:1nhop}
\end{figure}

\begin{table}
  \caption{On-site energies and nearest neighbor Mn-O and O-O hopping
    amplitudes obtained for the $d$-$p$ MLWFs in the cubic structure
    with FM order. In addition, the hopping between $t_{2g}$ orbitals
    along the shortest Mn-Mn distance is also listed. All values are
    given in eV.}
  \label{tab:hamme}
  \newcolumntype{.}{D{.}{.}{3.3}}
  \begin{ruledtabular}
    \begin{tabular*}{\columnwidth}{l@{\extracolsep{\fill}}..}
      &\multicolumn{1}{r}{majority spin}
      &\multicolumn{1}{r}{minority spin}\\
      \hline
      $\varepsilon[\text{Mn}(e_g)]$ & 12.459 & 16.074 \\
      $\varepsilon[\text{Mn}(t_{2g})]$ & 11.031 & 15.104 \\
      $\varepsilon[\text{O}(p_\sigma)]$ & 9.400 & 9.603 \\
      $\varepsilon[\text{O}(p_\pi)]$ & 9.888 & 10.432 \\
      \hline
      $t_{eg,p\sigma}$ & 1.768 & 1.861 \\
      $t_{t2g,p\pi}$ & -0.852 & -0.982 \\
      \hline
      $t_{p\sigma,p\sigma}$ & -0.444 & -0.409 \\
      $t_{p\sigma,p\pi}$ & 0.306 & 0.265 \\
      $t_{p\pi,p\pi}^{\perp}$ & -0.337 & -0.304 \\
      $t_{p\pi,p\pi}^{\parallel}$ & -0.014 & -0.092 \\
      \hline
      $t_{t2g,t2g}$ & -0.069 & -0.105 \\
    \end{tabular*}
  \end{ruledtabular}
\end{table}

\begin{figure}
  \includegraphics[width=\columnwidth]{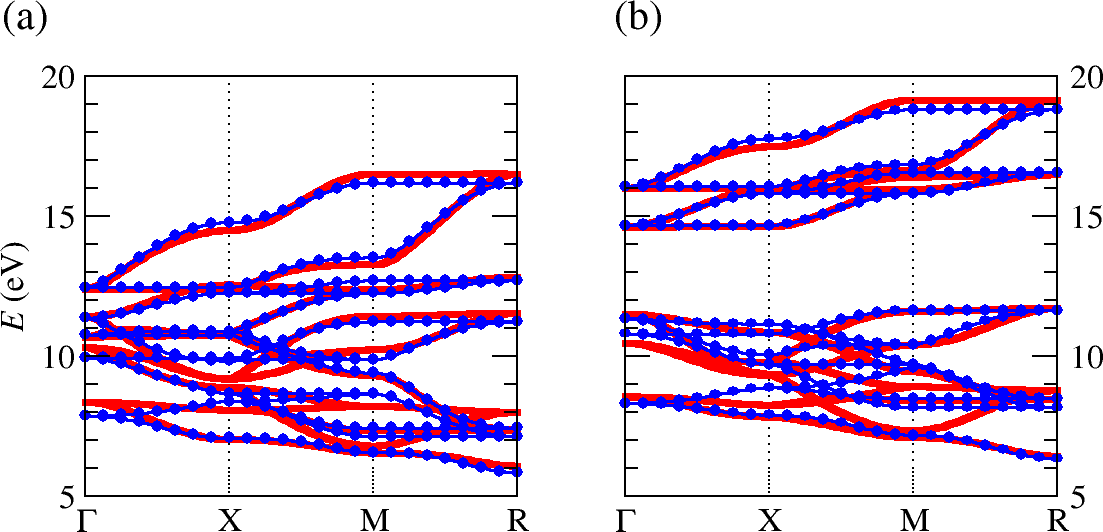}
  \caption{(Color online) Band dispersion calculated from the MLWFs
    (thick line) and from the truncated TB model described in the main
    text (dots and line); (a): majority spin, (b): minority spin.}
  \label{fig:TB-bands}
\end{figure}

In this appendix we give a more quantitative account of the $d$-$p$ TB
parametrization obtained from the MLWFs for simple cubic LaMnO$_{3}$
and FM order.

There are 2 different hopping parameters between nearest neighbor
Mn($d$) and O($p$) states ($t_{eg,p\sigma}$, $t_{t2g,p\pi}$) and 4
independent hopping parameters between closest neighbor O($p$) states
($t_{p\sigma,p\sigma}$, $t_{p\sigma,p\pi}$, $t_{p\pi,p\pi}^{\perp}$,
$t_{p\pi,p\pi}^{\parallel}$) that are allowed by the symmetry of the
system. All of these hopping parameters are illustrated in
Fig.~\ref{fig:1nhop}. The corresponding values together with the
various on-site energies are summarized in Table~\ref{tab:hamme}.  It
can be seen that the minority spin hopping amplitudes corresponding to
nearest neighbor $d$-$p$ hopping are slightly larger than for majority
spin, whereas the opposite is the case for the closest neighbor
$p$-$p$ hoppings (except for $t_{p\pi,p\pi}^{\parallel}$).

Within the two-center approximation to the linear combination of
atomic orbitals (LCAO) method all closest neighbor O($p$)-O($p$)
hoppings can be expressed as linear combination of purely $\sigma$-
and $\pi$-type hopping parameters, $t(pp\sigma)$ and
$t(pp\pi)$.~\cite{1954_slater-koster} Thereby, $t(pp\pi)$ corresponds
directly to $t_{p\pi,p\pi}^{\parallel}$ in our notation. Calculating
an estimate for $t(pp\sigma)$ from each of the remaining MLWF hopping
parameters as $(t_{p\pi,p\pi}^{\parallel}-2t_{p\sigma,p\sigma})$,
$(t_{p\pi,p\pi}^{\parallel}-2t_{p\pi,p\pi}^{\perp})$, or
$(-t_{p\pi,p\pi}^{\parallel}+2t_{p\sigma,p\pi})$, we obtain values of
0.873/0.725~eV, 0.660/0.516~eV, or 0.627/0.622~eV, for
majority/minority spin, respectively. The spread in these values shows
to what extent the approximation of rigid atomic orbitals is not
fulfilled within the $d$-$p$ set of MLWFs.

In a previous study, LCAO parameters for manganites were deduced based
on a cluster-model analysis of photoemission
spectra,~\cite{1996_mizokawa} leading to the following values for the
hopping integrals: $t(pd\sigma)=1.8$~eV, $t(pd\pi)=0.9$~eV,
$t(pp\sigma)=0.60$~eV, $t(pp\pi)=-0.15$~eV. These values are in
reasonable agreement with the values obtained from the MLWFs listed in
Table~\ref{tab:hamme}.

In Figure~\ref{fig:1nhop} we show a comparison of the dispersion
between the MLWF bands and a 14-band TB model calculated using the
parameters given in Table~\ref{tab:hamme}, i.e. by neglecting all
other further neighbor hoppings. The overall band dispersion is rather
well reproduced in the truncated TB model. However, due to a rather
slow decay of hopping amplitudes with further neighbor distance,
certain features are not reproduced. In addition, neglecting
further-neighbor hoppings also leads to a slight underestimation of
the total bandwidth in the truncated TB model. We find that there are
overall 14 different further neighbor hoppings with magnitudes in the
range of (0.020-0.113)~eV. Out of these only the direct hopping
between neighboring $t_{2g}$ orbitals (last line in
Table~\ref{tab:hamme}) leads to a significant improvement of the TB
bands and is therefore included in Table~\ref{tab:hamme}.

\bibliography{references}

\begin{thebibliography}{40}
\expandafter\ifx\csname natexlab\endcsname\relax\def\natexlab#1{#1}\fi
\expandafter\ifx\csname bibnamefont\endcsname\relax
  \def\bibnamefont#1{#1}\fi
\expandafter\ifx\csname bibfnamefont\endcsname\relax
  \def\bibfnamefont#1{#1}\fi
\expandafter\ifx\csname citenamefont\endcsname\relax
  \def\citenamefont#1{#1}\fi
\expandafter\ifx\csname url\endcsname\relax
  \def\url#1{\texttt{#1}}\fi
\expandafter\ifx\csname urlprefix\endcsname\relax\def\urlprefix{URL }\fi
\providecommand{\bibinfo}[2]{#2}
\providecommand{\eprint}[2][]{\url{#2}}

\bibitem[{\citenamefont{Dagotto}(1994)}]{Dagotto:1994}
\bibinfo{author}{\bibfnamefont{E.}~\bibnamefont{Dagotto}},
  \bibinfo{journal}{Rev. Mod. Phys.} \textbf{\bibinfo{volume}{66}},
  \bibinfo{pages}{763} (\bibinfo{year}{1994}).

\bibitem[{\citenamefont{Imada et~al.}(1998)\citenamefont{Imada, Fujimori, and
  Tokura}}]{Imada/Fujimori/Tokura:1998}
\bibinfo{author}{\bibfnamefont{M.}~\bibnamefont{Imada}},
  \bibinfo{author}{\bibfnamefont{A.}~\bibnamefont{Fujimori}}, \bibnamefont{and}
  \bibinfo{author}{\bibfnamefont{Y.}~\bibnamefont{Tokura}},
  \bibinfo{journal}{Rev. Mod. Phys.} \textbf{\bibinfo{volume}{70}},
  \bibinfo{pages}{1039} (\bibinfo{year}{1998}).

\bibitem[{\citenamefont{Dagotto et~al.}(2001)\citenamefont{Dagotto, Hotta, and
  Moreo}}]{Dagotto/Hotta/Moreo:2001}
\bibinfo{author}{\bibfnamefont{E.}~\bibnamefont{Dagotto}},
  \bibinfo{author}{\bibfnamefont{T.}~\bibnamefont{Hotta}}, \bibnamefont{and}
  \bibinfo{author}{\bibfnamefont{A.}~\bibnamefont{Moreo}},
  \bibinfo{journal}{Phys. Rep.} \textbf{\bibinfo{volume}{344}},
  \bibinfo{pages}{1} (\bibinfo{year}{2001}).

\bibitem[{\citenamefont{Hohenberg and Kohn}(1964)}]{Hohenberg/Kohn:1964}
\bibinfo{author}{\bibfnamefont{P.}~\bibnamefont{Hohenberg}} \bibnamefont{and}
  \bibinfo{author}{\bibfnamefont{W.}~\bibnamefont{Kohn}},
  \bibinfo{journal}{Phys. Rev.} \textbf{\bibinfo{volume}{136}},
  \bibinfo{pages}{B864} (\bibinfo{year}{1964}).

\bibitem[{\citenamefont{Kohn and Sham}(1965)}]{Kohn/Sham:1965}
\bibinfo{author}{\bibfnamefont{W.}~\bibnamefont{Kohn}} \bibnamefont{and}
  \bibinfo{author}{\bibfnamefont{L.~J.} \bibnamefont{Sham}},
  \bibinfo{journal}{Phys. Rev.} \textbf{\bibinfo{volume}{140}},
  \bibinfo{pages}{A1133} (\bibinfo{year}{1965}).

\bibitem[{\citenamefont{Gunnarsson et~al.}(1989)\citenamefont{Gunnarsson,
  Andersen, Jepsen, and Zaanen}}]{Gunnarsson_et_al:1989}
\bibinfo{author}{\bibfnamefont{O.}~\bibnamefont{Gunnarsson}},
  \bibinfo{author}{\bibfnamefont{O.~K.} \bibnamefont{Andersen}},
  \bibinfo{author}{\bibfnamefont{O.}~\bibnamefont{Jepsen}}, \bibnamefont{and}
  \bibinfo{author}{\bibfnamefont{J.}~\bibnamefont{Zaanen}},
  \bibinfo{journal}{Phys. Rev. B} \textbf{\bibinfo{volume}{39}},
  \bibinfo{pages}{1708} (\bibinfo{year}{1989}).

\bibitem[{\citenamefont{Hybertsen et~al.}(1989)\citenamefont{Hybertsen,
  Schl{\"{u}}ter, and Christensen}}]{Hybertsen/Schlueter/Christensen:1989}
\bibinfo{author}{\bibfnamefont{M.~S.} \bibnamefont{Hybertsen}},
  \bibinfo{author}{\bibfnamefont{M.}~\bibnamefont{Schl{\"{u}}ter}},
  \bibnamefont{and} \bibinfo{author}{\bibfnamefont{N.~E.}
  \bibnamefont{Christensen}}, \bibinfo{journal}{Phys. Rev. B}
  \textbf{\bibinfo{volume}{39}}, \bibinfo{pages}{9028} (\bibinfo{year}{1989}).

\bibitem[{\citenamefont{Ederer et~al.}(2007)\citenamefont{Ederer, Lin, and
  Millis}}]{Ederer/Lin/Millis:2007}
\bibinfo{author}{\bibfnamefont{C.}~\bibnamefont{Ederer}},
  \bibinfo{author}{\bibfnamefont{C.}~\bibnamefont{Lin}}, \bibnamefont{and}
  \bibinfo{author}{\bibfnamefont{A.~J.} \bibnamefont{Millis}},
  \bibinfo{journal}{Phys. Rev. B} \textbf{\bibinfo{volume}{76}},
  \bibinfo{pages}{155105} (\bibinfo{year}{2007}).

\bibitem[{\citenamefont{Kov\'a\v{c}ik and Ederer}(2010)}]{Kovacik/Ederer:2010}
\bibinfo{author}{\bibfnamefont{R.}~\bibnamefont{Kov\'a\v{c}ik}}
  \bibnamefont{and} \bibinfo{author}{\bibfnamefont{C.}~\bibnamefont{Ederer}},
  \bibinfo{journal}{Phys. Rev. B} \textbf{\bibinfo{volume}{81}},
  \bibinfo{pages}{245108} (\bibinfo{year}{2010}).

\bibitem[{\citenamefont{M{\"{u}}ller et~al.}(1998)\citenamefont{M{\"{u}}ller,
  Anisimov, Rice, Dasgupta, and Saha-Dasgupta}}]{Mueller_et_al:1998}
\bibinfo{author}{\bibfnamefont{T.~F.~A.} \bibnamefont{M{\"{u}}ller}},
  \bibinfo{author}{\bibfnamefont{V.}~\bibnamefont{Anisimov}},
  \bibinfo{author}{\bibfnamefont{T.~M.} \bibnamefont{Rice}},
  \bibinfo{author}{\bibfnamefont{I.}~\bibnamefont{Dasgupta}}, \bibnamefont{and}
  \bibinfo{author}{\bibfnamefont{T.}~\bibnamefont{Saha-Dasgupta}},
  \bibinfo{journal}{Phys. Rev. B} \textbf{\bibinfo{volume}{57}},
  (\bibinfo{year}{1998}).

\bibitem[{\citenamefont{Ku et~al.}(2002)\citenamefont{Ku, Rosner, Pickett, and
  Scalettar}}]{Ku_et_al:2002}
\bibinfo{author}{\bibfnamefont{W.}~\bibnamefont{Ku}},
  \bibinfo{author}{\bibfnamefont{H.}~\bibnamefont{Rosner}},
  \bibinfo{author}{\bibfnamefont{W.~E.} \bibnamefont{Pickett}},
  \bibnamefont{and} \bibinfo{author}{\bibfnamefont{R.~T.}
  \bibnamefont{Scalettar}}, \bibinfo{journal}{Phys. Rev. Lett.}
  \textbf{\bibinfo{volume}{89}}, \bibinfo{pages}{167204}
  (\bibinfo{year}{2002}).

\bibitem[{\citenamefont{Zurek et~al.}(2005)\citenamefont{Zurek, Jepsen, and
  Andersen}}]{Zurek/Jepsen/Andersen:2005}
\bibinfo{author}{\bibfnamefont{E.}~\bibnamefont{Zurek}},
  \bibinfo{author}{\bibfnamefont{O.}~\bibnamefont{Jepsen}}, \bibnamefont{and}
  \bibinfo{author}{\bibfnamefont{O.~K.} \bibnamefont{Andersen}},
  \bibinfo{journal}{ChemPhysChem} \textbf{\bibinfo{volume}{6}},
  \bibinfo{pages}{1934} (\bibinfo{year}{2005}).

\bibitem[{\citenamefont{Lechermann et~al.}(2006)\citenamefont{Lechermann,
  Georges, Poteryaev, Biermann, Posternak, Yamasaki, and
  Andersen}}]{Lechermann_et_al:2006}
\bibinfo{author}{\bibfnamefont{F.}~\bibnamefont{Lechermann}},
  \bibinfo{author}{\bibfnamefont{A.}~\bibnamefont{Georges}},
  \bibinfo{author}{\bibfnamefont{A.}~\bibnamefont{Poteryaev}},
  \bibinfo{author}{\bibfnamefont{S.}~\bibnamefont{Biermann}},
  \bibinfo{author}{\bibfnamefont{M.}~\bibnamefont{Posternak}},
  \bibinfo{author}{\bibfnamefont{A.}~\bibnamefont{Yamasaki}}, \bibnamefont{and}
  \bibinfo{author}{\bibfnamefont{O.~K.} \bibnamefont{Andersen}},
  \bibinfo{journal}{Phys. Rev. B} \textbf{\bibinfo{volume}{74}},
  \bibinfo{pages}{125120} (\bibinfo{year}{2006}).

\bibitem[{\citenamefont{Solovyev}(2006)}]{Solovyev:2006}
\bibinfo{author}{\bibfnamefont{I.~V.} \bibnamefont{Solovyev}},
  \bibinfo{journal}{Phys. Rev. B} \textbf{\bibinfo{volume}{73}},
  \bibinfo{pages}{155117} (\bibinfo{year}{2006}).

\bibitem[{\citenamefont{Perdew and Zunger}(1981)}]{Perdew/Zunger:1981}
\bibinfo{author}{\bibfnamefont{J.~P.} \bibnamefont{Perdew}} \bibnamefont{and}
  \bibinfo{author}{\bibfnamefont{A.}~\bibnamefont{Zunger}},
  \bibinfo{journal}{Phys. Rev. B} \textbf{\bibinfo{volume}{23}},
  \bibinfo{pages}{5048} (\bibinfo{year}{1981}).

\bibitem[{\citenamefont{Perdew et~al.}(1996)\citenamefont{Perdew, Burke, and
  Ernzerhof}}]{1996_perdew}
\bibinfo{author}{\bibfnamefont{J.~P.} \bibnamefont{Perdew}},
  \bibinfo{author}{\bibfnamefont{K.}~\bibnamefont{Burke}}, \bibnamefont{and}
  \bibinfo{author}{\bibfnamefont{M.}~\bibnamefont{Ernzerhof}},
  \bibinfo{journal}{Phys. Rev. Lett.} \textbf{\bibinfo{volume}{77}},
  \bibinfo{pages}{3865} (\bibinfo{year}{1996}).

\bibitem[{\citenamefont{Liechtenstein et~al.}(1995)\citenamefont{Liechtenstein,
  Anisimov, and Zaanen}}]{Liechtenstein/Anisimov/Zaanen:1995}
\bibinfo{author}{\bibfnamefont{A.~I.} \bibnamefont{Liechtenstein}},
  \bibinfo{author}{\bibfnamefont{V.~I.} \bibnamefont{Anisimov}},
  \bibnamefont{and} \bibinfo{author}{\bibfnamefont{J.}~\bibnamefont{Zaanen}},
  \bibinfo{journal}{Phys. Rev. B} \textbf{\bibinfo{volume}{52}},
  (\bibinfo{year}{1995}).

\bibitem[{\citenamefont{Dudarev et~al.}(1998)\citenamefont{Dudarev, Botton,
  Savrasov, Humphreys, and Sutton}}]{1998_dudarev}
\bibinfo{author}{\bibfnamefont{S.~L.} \bibnamefont{Dudarev}},
  \bibinfo{author}{\bibfnamefont{G.~A.} \bibnamefont{Botton}},
  \bibinfo{author}{\bibfnamefont{S.~Y.} \bibnamefont{Savrasov}},
  \bibinfo{author}{\bibfnamefont{C.~J.} \bibnamefont{Humphreys}},
  \bibnamefont{and} \bibinfo{author}{\bibfnamefont{A.~P.}
  \bibnamefont{Sutton}}, \bibinfo{journal}{Phys. Rev. B}
  \textbf{\bibinfo{volume}{57}}, \bibinfo{pages}{1505} (\bibinfo{year}{1998}).

\bibitem[{\citenamefont{Becke}(1993)}]{Becke:1993}
\bibinfo{author}{\bibfnamefont{A.~D.} \bibnamefont{Becke}},
  \bibinfo{journal}{J.~Chem. Phys.} \textbf{\bibinfo{volume}{98}},
  \bibinfo{pages}{1372} (\bibinfo{year}{1993}).

\bibitem[{\citenamefont{Marzari and Vanderbilt}(1997)}]{1997_marzari}
\bibinfo{author}{\bibfnamefont{N.}~\bibnamefont{Marzari}} \bibnamefont{and}
  \bibinfo{author}{\bibfnamefont{D.}~\bibnamefont{Vanderbilt}},
  \bibinfo{journal}{Phys.~Rev.~B} \textbf{\bibinfo{volume}{56}},
  \bibinfo{pages}{12847} (\bibinfo{year}{1997}).

\bibitem[{\citenamefont{Souza et~al.}(2001)\citenamefont{Souza, Marzari, and
  Vanderbilt}}]{2001_souza}
\bibinfo{author}{\bibfnamefont{I.}~\bibnamefont{Souza}},
  \bibinfo{author}{\bibfnamefont{N.}~\bibnamefont{Marzari}}, \bibnamefont{and}
  \bibinfo{author}{\bibfnamefont{D.}~\bibnamefont{Vanderbilt}},
  \bibinfo{journal}{Phys.~Rev.~B} \textbf{\bibinfo{volume}{65}},
  \bibinfo{pages}{035109} (\bibinfo{year}{2001}).

\bibitem[{\citenamefont{Mostofi et~al.}(2008)\citenamefont{Mostofi, Yates, Lee,
  Souza, Vanderbilt, and Marzari}}]{Mostofi_et_al:2008}
\bibinfo{author}{\bibfnamefont{A.~A.} \bibnamefont{Mostofi}},
  \bibinfo{author}{\bibfnamefont{J.~R.} \bibnamefont{Yates}},
  \bibinfo{author}{\bibfnamefont{Y.-S.} \bibnamefont{Lee}},
  \bibinfo{author}{\bibfnamefont{I.}~\bibnamefont{Souza}},
  \bibinfo{author}{\bibfnamefont{D.}~\bibnamefont{Vanderbilt}},
  \bibnamefont{and} \bibinfo{author}{\bibfnamefont{N.}~\bibnamefont{Marzari}},
  \bibinfo{journal}{Comp. Phys. Comm.} \textbf{\bibinfo{volume}{178}},
  \bibinfo{pages}{685} (\bibinfo{year}{2008}).

\bibitem[{\citenamefont{Coey et~al.}(1999)\citenamefont{Coey, Viret, and
  Moln{\'{a}}r}}]{Coey/Viret/Molnar:1999}
\bibinfo{author}{\bibfnamefont{J.~M.~D.} \bibnamefont{Coey}},
  \bibinfo{author}{\bibfnamefont{M.}~\bibnamefont{Viret}}, \bibnamefont{and}
  \bibinfo{author}{\bibfnamefont{S.~v.} \bibnamefont{Moln{\'{a}}r}},
  \bibinfo{journal}{Adv. Phys.} \textbf{\bibinfo{volume}{48}},
  \bibinfo{pages}{167} (\bibinfo{year}{1999}).

\bibitem[{\citenamefont{Elemans et~al.}(1971)\citenamefont{Elemans, Laar, Veen,
  and Loopstra}}]{Elemans_et_al:1971}
\bibinfo{author}{\bibfnamefont{J.~B. A.~A.} \bibnamefont{Elemans}},
  \bibinfo{author}{\bibfnamefont{B.~V.} \bibnamefont{Laar}},
  \bibinfo{author}{\bibfnamefont{K.~R. V.~d.} \bibnamefont{Veen}},
  \bibnamefont{and} \bibinfo{author}{\bibfnamefont{B.~O.}
  \bibnamefont{Loopstra}}, \bibinfo{journal}{J. Solid State Chemistry}
  \textbf{\bibinfo{volume}{3}}, \bibinfo{pages}{238} (\bibinfo{year}{1971}).

\bibitem[{\citenamefont{Pickett and Singh}(1996)}]{Pickett/Singh:1996}
\bibinfo{author}{\bibfnamefont{W.~E.} \bibnamefont{Pickett}} \bibnamefont{and}
  \bibinfo{author}{\bibfnamefont{D.~J.} \bibnamefont{Singh}},
  \bibinfo{journal}{Phys. Rev. B} \textbf{\bibinfo{volume}{53}},
  \bibinfo{pages}{1146} (\bibinfo{year}{1996}).

\bibitem[{\citenamefont{Satpathy et~al.}(1996)\citenamefont{Satpathy,
  Popovi{\'c}, and Vukajlovi{\'c}}}]{Satpathy/Popovic/Vukajlovic:1996}
\bibinfo{author}{\bibfnamefont{S.}~\bibnamefont{Satpathy}},
  \bibinfo{author}{\bibfnamefont{Z.~S.} \bibnamefont{Popovi{\'c}}},
  \bibnamefont{and} \bibinfo{author}{\bibfnamefont{F.~R.}
  \bibnamefont{Vukajlovi{\'c}}}, \textbf{\bibinfo{volume}{76}},
  \bibinfo{pages}{960} (\bibinfo{year}{1996}).

\bibitem[{\citenamefont{Sawada et~al.}(1997)\citenamefont{Sawada, Morikawa,
  Terakura, and Hamada}}]{Sawada_et_al:1997}
\bibinfo{author}{\bibfnamefont{H.}~\bibnamefont{Sawada}},
  \bibinfo{author}{\bibfnamefont{Y.}~\bibnamefont{Morikawa}},
  \bibinfo{author}{\bibfnamefont{K.}~\bibnamefont{Terakura}}, \bibnamefont{and}
  \bibinfo{author}{\bibfnamefont{N.}~\bibnamefont{Hamada}},
  \bibinfo{journal}{Phys. Rev. B} \textbf{\bibinfo{volume}{56}},
  \bibinfo{pages}{12154} (\bibinfo{year}{1997}).

\bibitem[{\citenamefont{Giannozzi et~al.}(2009)\citenamefont{Giannozzi, Baroni,
  Bonini, Calandra, Car, Cavazzoni, Ceresoli, Chiarotti, Cococcioni, Dabo
  et~al.}}]{quantum-espresso}
\bibinfo{author}{\bibfnamefont{P.}~\bibnamefont{Giannozzi}},
  \bibinfo{author}{\bibfnamefont{S.}~\bibnamefont{Baroni}},
  \bibinfo{author}{\bibfnamefont{N.}~\bibnamefont{Bonini}},
  \bibinfo{author}{\bibfnamefont{M.}~\bibnamefont{Calandra}},
  \bibinfo{author}{\bibfnamefont{R.}~\bibnamefont{Car}},
  \bibinfo{author}{\bibfnamefont{C.}~\bibnamefont{Cavazzoni}},
  \bibinfo{author}{\bibfnamefont{D.}~\bibnamefont{Ceresoli}},
  \bibinfo{author}{\bibfnamefont{G.~L.} \bibnamefont{Chiarotti}},
  \bibinfo{author}{\bibfnamefont{M.}~\bibnamefont{Cococcioni}},
  \bibinfo{author}{\bibfnamefont{I.}~\bibnamefont{Dabo}}, \bibnamefont{et~al.},
  \bibinfo{journal}{J. Phys.: Condens. Matter} \textbf{\bibinfo{volume}{21}},
  \bibinfo{pages}{395502} (\bibinfo{year}{2009}).

\bibitem[{\citenamefont{Vanderbilt}(1990)}]{1990_vanderbilt}
\bibinfo{author}{\bibfnamefont{D.}~\bibnamefont{Vanderbilt}},
  \bibinfo{journal}{Phys.~Rev.~B} \textbf{\bibinfo{volume}{41}},
  \bibinfo{pages}{7892} (\bibinfo{year}{1990}).

\bibitem[{\citenamefont{Norby et~al.}(1995)\citenamefont{Norby, Andersen,
  Andersen, and Andersen}}]{1995_norby}
\bibinfo{author}{\bibfnamefont{P.}~\bibnamefont{Norby}},
  \bibinfo{author}{\bibfnamefont{I.~K.} \bibnamefont{Andersen}},
  \bibinfo{author}{\bibfnamefont{E.~K.} \bibnamefont{Andersen}},
  \bibnamefont{and} \bibinfo{author}{\bibfnamefont{N.}~\bibnamefont{Andersen}},
  \bibinfo{journal}{J.~Solid~State~Chem.} \textbf{\bibinfo{volume}{119}},
  \bibinfo{pages}{191} (\bibinfo{year}{1995}).

\bibitem[{\citenamefont{Kokalj}(2003)}]{2003_kokalj}
\bibinfo{author}{\bibfnamefont{A.}~\bibnamefont{Kokalj}},
  \bibinfo{journal}{Comp. Mater. Sci.} \textbf{\bibinfo{volume}{28}},
  \bibinfo{pages}{155 } (\bibinfo{year}{2003}), \bibinfo{note}{code available
  from http://www.xcrysden.org/.}

\bibitem[{\citenamefont{Kanamori}(1960)}]{Kanamori:1960}
\bibinfo{author}{\bibfnamefont{J.}~\bibnamefont{Kanamori}},
  \bibinfo{journal}{J. Appl. Phys. (Suppl.)} \textbf{\bibinfo{volume}{31}},
  \bibinfo{pages}{14} (\bibinfo{year}{1960}).

\bibitem[{foo({\natexlab{a}})}]{footnote:spin-index}
\bibinfo{note}{For simplicity of notation we suppress the spin index $\sigma$
  here. It should be understood that in general the eigenvalues
  $\epsilon_{l\mathbf{k}}$, the unitary matrices $\mathbf{U}^{(\mathbf{k})}$,
  and the occupation matrix of the MLWFs can be spin-dependent.}

\bibitem[{foo({\natexlab{b}})}]{footnote:j0}
\bibinfo{note}{We point out that Eq.~(\ref{eq:ldau}) corresponds to a
  simplified Hubbard interaction with exchange parameter $J=0$. While for a
  more general interaction with $J \neq 0$ the potential shifts would be
  different, our general conclusion would be unaffected. For simplicity of
  presentation we therefore only discuss the case $J=0$.}

\bibitem[{\citenamefont{Aichhorn et~al.}(2009)\citenamefont{Aichhorn,
  Pourovskii, Vildosola, Ferrero, Parcollet, Miyake, Georges, and
  Biermann}}]{Aichhorn_et_al:2009}
\bibinfo{author}{\bibfnamefont{M.}~\bibnamefont{Aichhorn}},
  \bibinfo{author}{\bibfnamefont{L.}~\bibnamefont{Pourovskii}},
  \bibinfo{author}{\bibfnamefont{V.}~\bibnamefont{Vildosola}},
  \bibinfo{author}{\bibfnamefont{M.}~\bibnamefont{Ferrero}},
  \bibinfo{author}{\bibfnamefont{O.}~\bibnamefont{Parcollet}},
  \bibinfo{author}{\bibfnamefont{T.}~\bibnamefont{Miyake}},
  \bibinfo{author}{\bibfnamefont{A.}~\bibnamefont{Georges}}, \bibnamefont{and}
  \bibinfo{author}{\bibfnamefont{S.}~\bibnamefont{Biermann}},
  \bibinfo{journal}{Phys. Rev. B} \textbf{\bibinfo{volume}{80}},
  \bibinfo{pages}{085101} (\bibinfo{year}{2009}).

\bibitem[{\citenamefont{Aryasetiawan and
  Gunnarsson}(1998)}]{Aryasetiawan/Gunnarsson:1998}
\bibinfo{author}{\bibfnamefont{F.}~\bibnamefont{Aryasetiawan}}
  \bibnamefont{and}
  \bibinfo{author}{\bibfnamefont{O.}~\bibnamefont{Gunnarsson}},
  \bibinfo{journal}{Rep. Prog. Phys.} \textbf{\bibinfo{volume}{61}},
  \bibinfo{pages}{237} (\bibinfo{year}{1998}).

\bibitem[{\citenamefont{Aryasetiawan et~al.}(2004)\citenamefont{Aryasetiawan,
  Imada, Georges, Kotliar, Biermann, and
  Lichtenstein}}]{Aryasetiawan_et_al:2004}
\bibinfo{author}{\bibfnamefont{F.}~\bibnamefont{Aryasetiawan}},
  \bibinfo{author}{\bibfnamefont{M.}~\bibnamefont{Imada}},
  \bibinfo{author}{\bibfnamefont{A.}~\bibnamefont{Georges}},
  \bibinfo{author}{\bibfnamefont{G.}~\bibnamefont{Kotliar}},
  \bibinfo{author}{\bibfnamefont{S.}~\bibnamefont{Biermann}}, \bibnamefont{and}
  \bibinfo{author}{\bibfnamefont{A.~I.} \bibnamefont{Lichtenstein}},
  \bibinfo{journal}{Phys. Rev. B} \textbf{\bibinfo{volume}{70}},
  \bibinfo{pages}{195104} (\bibinfo{year}{2004}).

\bibitem[{\citenamefont{Miyake and
  Aryasetiawan}(2008)}]{Miyake/Aryasetiawan:2008}
\bibinfo{author}{\bibfnamefont{T.}~\bibnamefont{Miyake}} \bibnamefont{and}
  \bibinfo{author}{\bibfnamefont{F.}~\bibnamefont{Aryasetiawan}},
  \bibinfo{journal}{Phys. Rev. B} \textbf{\bibinfo{volume}{77}},
  \bibinfo{pages}{085122} (\bibinfo{year}{2008}).

\bibitem[{\citenamefont{Slater and Koster}(1954)}]{1954_slater-koster}
\bibinfo{author}{\bibfnamefont{J.~C.} \bibnamefont{Slater}} \bibnamefont{and}
  \bibinfo{author}{\bibfnamefont{G.~F.} \bibnamefont{Koster}},
  \bibinfo{journal}{Phys. Rev.} \textbf{\bibinfo{volume}{94}},
  \bibinfo{pages}{1498} (\bibinfo{year}{1954}).

\bibitem[{\citenamefont{Mizokawa and Fujimori}(1996)}]{1996_mizokawa}
\bibinfo{author}{\bibfnamefont{T.}~\bibnamefont{Mizokawa}} \bibnamefont{and}
  \bibinfo{author}{\bibfnamefont{A.}~\bibnamefont{Fujimori}},
  \bibinfo{journal}{Phys. Rev. B} \textbf{\bibinfo{volume}{54}},
  \bibinfo{pages}{5368} (\bibinfo{year}{1996}).

\end{thebibliography}

\end{document}